\let\AAoriginalenddocument\enddocument
\documentclass{aa}
\let\enddocument\AAoriginalenddocument

\usepackage{graphicx}
\usepackage{txfonts}
\usepackage{xcolor}
\usepackage{multicol}
\usepackage{multirow}
\usepackage[utf8]{inputenc}
\usepackage{booktabs}
\usepackage{natbib}
\usepackage{ulem}

\bibpunct{(}{)}{;}{a}{}{,}

\usepackage[
    colorlinks=true,
    urlcolor=blue,
    citecolor=blue,
    linkcolor=blue
]{hyperref}

\usepackage{lastpage}

\makeatletter
\renewcommand*{\aa@pageof}{, page \thepage{} of \pageref*{LastPage}}
\makeatother

\begin{document}

   \title{The AIDA-TNG project: Galactic bar pattern speeds in self-interacting dark matter}
   
   \author{
Asiyeh Habibi\inst{1}\thanks{
\email{asiyeh.habibi@alumni.um.ac.ir}}
\and Mahmood Roshan\inst{1,2}
\and Giulia Despali\inst{3,6,7}
\and Mark Vogelsberger\inst{4}
\and Amir Ghari\inst{1,5}
\and Lauro Moscardini\inst{3,6,7}
}

\institute{
Department of Physics, Faculty of Science, Ferdowsi University of Mashhad,
P.O. Box 1436, Mashhad, Iran
\and Research Center for Modeling and Computational Sciences, Faculty of Science, Ferdowsi University of Mashhad, Mashhad, Iran
\and
Dipartimento di Fisica e Astronomia ``Augusto Righi'',
Alma Mater Studiorum Università di Bologna,
Via Gobetti 93/2, I-40129 Bologna, Italy
\and
Department of Physics, Massachusetts Institute of Technology,
77 Massachusetts Avenue, Cambridge, MA 02139, USA
\and
Department of Physics, Institute for Advanced Studies in Basic Sciences (IASBS), P.O. Box 11365-9161, Zanjan, Iran 
\and INAF-Osservatorio di Astrofisica e Scienza dello Spazio di Bologna, Via Piero Gobetti 93/3, I-40129 Bologna, Italy 
\and INFN-Sezione di Bologna, Viale Berti Pichat 6/2, I-40127 Bologna, Italy 
}

   \date{\today}

\abstract
{Cosmological simulations within the standard cold dark matter (CDM) paradigm systematically predict higher values of the dimensionless bar-speed parameter, $\mathcal{R} \equiv R_{\rm cr}/R_{\rm bar}$ (where $R_{\rm cr}$ and $R_{\rm bar}$ denote the corotation radius and the bar radius, respectively), than those inferred from observations of local galaxies, a discrepancy known as the fast-bar tension. Since bar slowdown is primarily driven by dynamical friction with the dark matter halo, bar dynamics is inherently sensitive to the properties of dark matter particles.}
{We aim to determine whether dark matter self-interactions alleviate or exacerbate the fast-bar tension, and to identify the physical mechanism behind the differences in bar dynamics between CDM and self-interacting dark matter (SIDM) models.}
{We use the $51.7\,{\rm Mpc}$ box of the AIDA-TNG cosmological simulations, run with three different dark matter models: standard CDM, and two SIMD models one with a constant cross-section model (SIDM1), and the other with a velocity-dependent model (vSIDM). We identify barred disk galaxies with stellar mass $M_* > 10^{10}\,M_{\odot}$ at redshifts $z=0.2,\,1.0,\,1.5,$ and $2.0$, and measure the bar fraction, bar strength, bar length, pattern speed $\Omega_{\rm p}$, and the ratio $\mathcal{R}=R_{\rm cr}/R_{\rm bar}$.}
{The pattern speeds in the SIDM1 and vSIDM models are generally lower than in CDM at nearly all redshifts probed in this study. The subsequent slowdown proceeds with nearly 10\% more effective in the SIDM models likely due to velocity-anisotropy. Consequently, at low redshift the mean value of $\mathcal{R}$ is significantly higher in the SIDM models, indicating that the fas-bar tension is further exacerbated in these scenarios.}
{Bars rotate more slowly in SIDM than in CDM, particularly in SIDM1. We show that the origin of this difference is closely tied to the epoch of bar formation rather than to dynamical friction alone.}

   \maketitle
%

\section{Introduction}
\label{intro}

Stellar bars are among the most common morphological features of disk galaxies, being present in roughly half to two-thirds of the disk galaxy population in the local Universe \citep{erwin2018}. Far from being passive structures, bars are principal drivers of the secular evolution of their hosts: they redistribute angular momentum within the disk--halo system, channel gas inflows toward galactic centers, and influence the growth of bulges and supermassive black holes \citep{2009ASSP....8..159G, math2025}. The bar pattern speed, $\Omega_{\rm p}$, which quantifies the angular velocity of the bar, is a fundamental parameter governing these processes: it sets the location of the resonances through which the bar couples to the other galactic components and regulates the exchange of angular momentum between the bar and the dark matter halo \citep{at1992, Athanassoula2013, 2014RvMP...86....1S}. Because this coupling depends directly on the properties of the surrounding halo, the pattern speeds of galactic bars provide a sensitive dynamical probe of the distribution, and potentially the fundamental nature, of dark matter.

To characterize the rotation rate of bars, the dimensionless parameter $\mathcal{R}$ is conveniently defined as the ratio of the corotation radius ($R_{\rm cr}$) to the bar radius ($R_{\rm bar}$), i.e., $\mathcal{R} \equiv R_{\rm cr}/R_{\rm bar}$. The corotation radius is related to the pattern speed $\Omega_{\rm p}$ by the equation $\Omega(R_{\rm cr})=\Omega_{\rm p}$, where $\Omega$ represents the angular velocity derived from the circular velocity of the stars, given by $v_{\rm cir}=R\,\Omega(R)$. Bars with $\mathcal{R} < 1.4$ are conventionally classified as ``fast'', while those with $\mathcal{R} > 1.4$ are ``slow''. Cosmological simulations such as EAGLE \citep{eagle} and IllustrisTNG \citep{tng1,tng2,tng3,tng4,tng5,TNG502,TNG501,TNG503}, where the dynamics and properties of stellar bars have been widely studied \citep[e.g.,][]{2022MNRAS.514.1006I,2024A&A...684A.179R,2025A&A...698A..20R,2025MNRAS.542.3154G,2026MNRAS.546ag263G,2026arXiv260912339C}, predict that most bars are slow, whereas observations of local galaxies indicate that a considerable fraction of bars are fast \citep{Aguerri2015, algorry2017, Cuomo2020, 2021MNRAS.508..926R}. For instance \cite{Geron2023} measured the pattern speed of 225 barred galaxies\footnote{While \cite{Geron2023} report pattern speed measurements for 225 barred galaxies, it is likely that only a subset of these have high-accuracy determinations of $\Omega_{\rm p}$. Notably, among the 50 randomly selected galaxies presented in their Table 3, more than 40\% of the measured pattern speeds exhibit relative errors exceeding 30\%.}. This study benefits from a larger sample of galaxies than previous studies. They found that about 62\% of the bars in their sample are slow. However, the mean value of $\mathcal{R}$ is approximately 1.7, which remains significantly different from the TNG50 value of about 3.1 \citep{2021MNRAS.508..926R}. This discrepancy is referred to as the ``fast-bar tension''. The only plausible resolution of this tension proposed to date is that of \citet{2022ApJ...940...61F}, who attribute the large values of $\mathcal{R}$ in TNG50 to the short-bar problem, that is, the absence of sufficiently long bars in the simulation, rather than to the pattern speed itself. They further argue that the short-bar problem may partly reflect inconsistencies in the comparison between simulations and observational data.

The physical origin of slow bars in simulations is well understood. A rotating bar embedded in a live dark matter halo loses angular momentum to the halo through resonant interactions, a process usually described as dynamical friction \citep{tw1984, weinberg1985, debattista2000, athanassoula2003}. As a result, bars in $N$-body simulations robustly decelerate with cosmic time, and their corotation radii migrate outward, driving $\mathcal{R}$ into the slow regime. The apparent reduction of the effect of this mechanism in observed galaxies therefore raises critical questions about the angular momentum exchange between bars and their surrounding halos, as well as the influence of baryonic processes on bar dynamics \citep{debattista2000, Fragkoudi2021}. One class of proposed solutions invokes baryonic physics: in simulations with high baryon dominance in the inner regions, dynamical friction is weakened and bars remain fast \citep{Fragkoudi2021}. An alternative possibility is that the discrepancy reflects the nature of dark matter itself, since the efficiency of dynamical friction depends on the microphysics of the dark matter. A notable example is ultralight (fuzzy) dark matter, for which the frictional drag can be substantially suppressed compared to the classical Chandrasekhar expectation \citep{hui2017, lancaster2020}. Different dark matter models therefore predict different frictional couplings between bars and halos, and hence potentially distinguishable distributions of bar pattern speeds.

Self-interacting dark matter (SIDM) is one of the most extensively studied alternatives to collisionless cold dark matter (CDM) \citep{spergel2000}; see \cite{tulin2018} and \cite{adhikari2022} for reviews. In this scenario, elastic scattering between dark matter particles transports heat through the halo and drives the inner regions toward a quasi-isothermal state. In dark-matter-only simulations, this thermalization can transform central cusps into approximately constant-density cores, motivating SIDM as a possible solution for the core--cusp problem \citep{deblok2010,Vogelsberger2012,Zavala2013} and the diversity of dwarf-galaxy rotation curves \citep{Creasey2017}. Velocity-dependent self-interactions can produce sizeable effects in low-velocity systems while remaining weaker at cluster scales where observational constraints are available \citep{Vogelsberger2012,Zavala2013}. More generally, the connection between dark-sector microphysics and structure formation can be formulated within effective frameworks such as ETHOS \citep{CyrRacine2016,Vogelsberger2016}, while extensions including inelastic self-interactions have also been explored \citep{Vogelsberger2019}. Gravothermal evolution can eventually drive some SIDM systems from an initial core toward a core-collapsed cusp; this process has been investigated in detail for subhalos \citep{Zeng2022}. The picture becomes considerably richer when baryons are also included: in the baryon-dominated regions of massive disk galaxies, self-interactions couple the dark matter to the deep baryonic potential, and the halo response can range from mild coring to contraction stronger than in CDM, occasionally producing inner slopes steeper than the CDM counterparts at Milky Way masses \citep{Creasey2017,Sameie2018,Rose2023,despali2026}. Since dynamical friction depends on the local dark matter density, velocity distribution, and anisotropy, these structural and kinematic differences are expected to leave a direct imprint on the evolution of bar pattern speeds.

Indeed, recent theoretical work indicates that dark matter self-interactions modify the processes that control bar evolution. \cite{fischer2024} showed, using idealized simulations of massive perturbers, that self-interactions typically enhance the dynamical friction force experienced by a supersonic perturber relative to the collisionless case. More recently, \cite{bosch2026} used high-resolution idealized $N$-body simulations and kinetic theory to demonstrate that bars in SIDM halos form earlier and grow stronger than in CDM, because self-interactions broaden the bar-halo resonances and thereby enhance the transfer of angular momentum from the disk to the halo; notably, they showed that this effect is not driven by core formation. These results, together with the growing number of strong bars detected by JWST at high redshift \citep{guo2023, leconte2024}, suggest that the statistics and dynamics of barred galaxies constitute a promising probe of dark matter self-interactions. However, all of these studies rely on idealized, non-cosmological setups, and the fast-bar tension has not yet been fully investigated in the context of SIDM. It therefore remains an open question whether self-interactions alleviate or exacerbate the discrepancy between simulated and observed bar pattern speeds. In this paper, we address this question using cosmological hydrodynamical simulations.

Addressing this question requires cosmological simulations that combine a realistic galaxy formation model with SIDM. Such simulations have recently become available: the AIDA-TNG project \citep{despali2025} evolves the same initial conditions and the fiducial IllustrisTNG galaxy formation model under CDM as well as several alternative dark matter scenarios, including a SIDM model with constant cross-section model with $\sigma/m = 1\,{\rm cm^2\,g^{-1}}$ (SIDM1) and a velocity-dependent SIDM model (vSIDM) in which the cross-section is large in dwarf-scale halos and decreases toward cluster scales. Because the different runs share identical initial conditions, individual galaxies can be matched across dark matter models, allowing a direct, object-by-object assessment of how dark matter microphysics affects bar dynamics, with cosmologically realistic assembly histories, gas inflows, and mergers. This matching is essential because bar formation and evolution are sensitive to each galaxy's assembly history, which otherwise would make it difficult to disentangle the systematic effects of the dark matter model from intrinsic galaxy-to-galaxy variations.

In this paper, we use the highest resolution box of AIDA-TNG to carry out the first measurement of bar pattern speeds in SIDM cosmological simulations. We identify barred disk galaxies with stellar mass $M_* > 10^{10}\,M_{\odot}$ in the CDM, SIDM1, and vSIDM runs at four redshifts $z = 0.2, 1.0, 1.5$, and $2.0$, and compare the bar fraction, bar strength, bar length, pattern speed, and the parameter $\mathcal{R}$ across the three models, both for the full populations and for the matched subsample. We then investigate whether the differences we find are consistent with a different efficiency of dynamical friction by comparing the halo density, velocity dispersion, and velocity anisotropy profiles of the matched galaxies, and whether the differences are already imprinted at the epoch of bar formation.

This paper is organized as follows. In Sect.~\ref{Data} we introduce the AIDA-TNG simulations and describe our methods for selecting disk galaxies, identifying bars, and measuring the bar length, pattern speed, and corotation radius. In Sect.~\ref{results} we present the bar properties in the three dark matter models, compare the matched galaxies across models, and discuss the role of dynamical friction in the observed differences. Finally, we summarize our findings and discuss their implications for the fast-bar tension in Sect.~\ref{conclusion}. Throughout this paper, we adopt cosmological parameters consistent with those used in the AIDA-TNG simulation: $H_{0} = 67.74\,\rm{km\,s^{-1}\,Mpc^{-1}}$, $\Omega_{\mathrm{b},0} = 0.0486$, $\Omega_{\mathrm{m},0} = 0.3089$, $\Omega_{\mathrm{\Lambda},0} = 0.6911$ \citep{despali2025}.

\section{Data and methods}
\label{Data}

\subsection{AIDA-TNG simulations}
\label{AIDA}

The AIDA-TNG simulations, Alternative Dark Matter in the TNG Universe (hereafter AIDA), comprise cosmological volumes with increasing resolution and decreasing size, each simulated under cold as well as warm and self-interacting dark matter models, both with and without baryonic physics. A full description of the simulation suite can be found in   \citet{despali2025}. In this paper, we focus on the higher resolution simulation with a box size of 51.7 Mpc and mass resolutions of $m_{\rm DM} = 3.6 \times 10^6~M_{\odot}$ for dark matter and $m_{\rm bar} = 6.8 \times 10^5~M_{\odot}$ for baryons. In terms of mass resolution, this run, referred to as 50/A in \cite{despali2025}, is comparable to the TNG100 simulation in IllustrisTNG. AIDA-TNG employs the TNG galaxy formation model, which updates upon the previous Illustris model \citep{Torrey14}, including a revised kinetic AGN feedback model for the low accretion state \citep{weinberger17} and an improved parametrisation of galactic winds \citep{pillepich18}. The model includes (i) a sub-resolution treatment of the interstellar medium (ISM) as a two-phase gas where cold clumps are embedded in a smooth, hot phase produced by supernova explosions \citep{springel03}; (ii) feedback from supernova explosions and stellar winds, in the form of kinetic and thermal energy; (iii) the production and evolution of nine elements (H, He, C, N, O, Ne, Mg, Si, and Fe), as well as the tracking of the overall gas metallicity and a model for Europium; (iv) density-, redshift-, metallicity-, and temperature-dependent cooling; and (v) super-massive black hole (SMBH) formation, growth and feedback in two different modes (quasar/thermal and kinetic/low-state).

We select barred galaxies with stellar masses $M_* > 10^{10.0}\, M_{\odot}$ from 50/A at redshifts $z=0.2, 1.0, 1.5$ and 2. This mass range is chosen to ensure sufficient resolution, or number of particles, for reliable identification of stellar bars. With this selection, each galaxy contains at least $10^4$ stellar particles. Although the CDM and vSIDM simulations were evolved down to redshift $z=0$, the SIDM1 simulation is ongoing and currently available until $z=0.18$. Consequently, we adopt $z=0.2$ as our baseline low-redshift limit, ensuring a uniform comparison across all three models. 

In the following subsections, we briefly describe our criteria for identifying galactic bars and the methods used to measure their properties. More details can be found in \cite{2021MNRAS.508..926R}.

\subsection{Selecting disk galaxies}
\label{sample-selection}
To identify barred galactic disks, we first need to distinguish disk galaxies within our sample. The center of each galaxy is assumed to coincide with the center of mass of its inner regions. By inner region, we mean the volume enclosed within a sphere of radius $4\,\text{kpc}$ centered on the center of mass of the entire stellar component of the galaxy. We begin by calculating the direction of the total angular momentum vector for stellar particles within the stellar half-mass radius\footnote{Since the center of the galaxy has been determined, we simply calculate the radius of the sphere that encloses half of the total stellar mass.}. We align the $z$-axis with this direction and then apply two criteria to select disk galaxies:
i) $k_{\rm{rot}} \geq 0.5$ and
ii) $F \leq 0.7$
Here, $k_{\rm{rot}}$ represents the fraction of rotational kinetic energy relative to total kinetic energy. Specifically, it is defined as the mass-weighted mean value of $v_{\phi}^2/v^2$ within 30 kpc, where $v$ is the total velocity and $v_{\phi}$ is the azimuthal velocity of each stellar particle \citep{sales2010}.
The morphological flatness parameter $F$ is defined as $F \equiv M_1/\sqrt{M_2 M_3}$, where $M_i$ are the eigenvalues of the moment of inertia tensor of the stellar particles, sorted such that $M_1 \leq M_2 \leq M_3$ \citep{genel2015}. These disk selection criteria have been adopted in a number of previous studies like \citep{2020ApJ...904..170Z, 2021MNRAS.508..926R, AHabibi}.

\subsection{Selecting barred disk galaxies}
\label{barred_disks}
By computing the $m = 2$ azimuthal Fourier component of the stellar mass distribution of the disk we define the bar strength. We project all of the particles to the $x-y$ plane, and divide the projected disk into annuli of fixed width $\Delta R=0.1$ kpc. The Fourier coefficients are given by:
\begin{eqnarray}
	a_{m} \left( R \right) &\equiv& \frac{1}{M \left( R \right)} \sum_{k=0}^{N} m_k \cos \left( m \phi_k \right), ~ {m} = 1, 2, .. \, , \\
	b_{m} \left( R \right) &\equiv& \frac{1}{M \left( R \right)} \sum_{k=0}^{N} m_k \sin \left( m \phi_k \right), ~ {m} = 1, 2, .. \, , 
\end{eqnarray}
where, $N$ represents the number of particles in the annulus, $R$ denotes the mean cylindrical radius of the annulus, $M$ is the total mass of particles within the annulus. Each particle is identified by the index $k$, with mass $m_k$ and azimuthal angle $\phi_k$. Now, the Fourier amplitude for mode $m$ at radius $R$ is defined as:
\begin{eqnarray}
    A_{m} \left( R \right) ~\equiv~ \sqrt{a_{m} \left( R \right)^2 + b_{m} \left( R \right)^2}, \, \,\,\,\, m\geq 1 .
\end{eqnarray}
Notice that with the definitions above we have $A_0=1$ \citep{ohta,Aguerri_2000,Athanassoula2013}. When an $m=2$ symmetric feature, such as a bar, is present in the disk, this function typically exhibits a pronounced maximum in the internal part of the disk. We use this maximum value to define the bar strength $A_2^{\text{max}} ~\equiv~ \max [A_2(R)]$ \citep{Laurikainen2008}. Bars are typically classified into two categories: strong bars, characterized by $A_2^{\text{max}} \geq 0.4$, and weak bars, defined by $0.2 \leq A_2^{\text{max}} < 0.4$. Conversely, whereas, disks with $A_2^{\text{max}} < 0.2$ are considered unbarred \citep{algorry2017}. We take into account both the weak and strong bars in our sample.

\subsection{Bar length measurement}
\label{bar-ln}

To measure the bar length we use the Fourier decomposition of the mass surface density presented in \cite{Aguerri_2000}. We first compute the intensity in the bar ($I_{\mathrm{b}}$) and inter-bar ($I_{\mathrm{ib}}$) zones, and find their ratio as a function of radius $R$ in the interval $(0.1,6)\,$kpc:
\begin{equation}
	\mathcal{I} \left( R \right) ~\equiv~ \frac{I_{\mathrm{b}} \left( R \right)}{I_{\mathrm{ib}} \left( R \right)} ~=~ \frac{A_0/2 ~+~ A_2 ~+~ A_4 ~+~ A_6}{A_0/2 ~-~ A_2 ~+~ A_4 ~-~ A_6 } \, .
	\label{Bar_interbar_ratio}
\end{equation}
The semi-major axis of the bar is defined as the outermost radius beyond which $\mathcal{I} \left( R \right)$ falls below $\left( \mathcal{I}^{\text{max}} + \mathcal{I}^{\text{min}} \right)/2$, where $\mathcal{I}^{\text{max}}$ and $\mathcal{I}^{\text{min}}$ are the maximum and minimum values of $\mathcal{I} \left( R \right)$, respectively. We refer to this semi-major axis as the bar radius, $R_{\rm bar}$; the bar length is $2R_{\rm bar}$. Throughout this paper, the rotation rate parameter $\mathcal{R}\equiv R_{\rm cr}/R_{\rm bar}$ is computed using $R_{\rm bar}$.
In some rare cases, it is necessary to extend the radial range beyond $6~\text{kpc}$ in order to measure the bar length. 
\begin{table*}[!t]
\caption{Galaxy counts at each redshift for the three dark matter models.}
\label{tab:galaxy_counts}
\centering
\begin{tabular}{l c c c c c}
\hline\hline
$z$ & $N_{\rm gal}$ & $N_{\rm disk}$ & $N_{\rm bar}$ & $N_{\rm rel}$ & $N_{\rm matched}$ \\
\hline
2.0 & $(274, 272, 304)$ & $(84, 90, 93)$    & $(40, 37, 31)$   & $(36, 24, 29)$ & 5  \\
1.5 & $(382, 384, 415)$ & $(163, 156, 167)$ & $(69, 65, 77)$   & $(58, 55, 65)$ & 19 \\
1.0 & $(509, 510, 549)$ & $(275, 275, 280)$ & $(96, 116, 125)$ & $(78, 90, 95)$ & 28 \\
0.2 & $(664, 662, 709)$ & $(413, 395, 417)$ & $(171, 164, 149)$ & $(155, 138, 127)$ & 46 \\
\hline
\end{tabular}
\tablefoot{$z$: redshift; $N_{\rm gal}$: number of galaxies with $M_* \geq 10^{10}\,M_\odot$ resolved with more than $10^4$ star particles; $N_{\rm disk}$: number of disk galaxies; $N_{\rm bar}$: number of barred galaxies; $N_{\rm rel}$: number of barred galaxies with reliable measurements of pattern speed, bar length, and corotation radius; $N_{\rm matched}$: number of matched barred systems at each redshift, i.e., galaxies that are barred and have reliable measurements of pattern speed, bar length, and corotation radius in all three models. Values in parentheses correspond to (CDM, vSIDM, SIDM1), respectively.}
\end{table*}

  \begin{figure}
  \centering
  \includegraphics[width=0.4\textwidth]{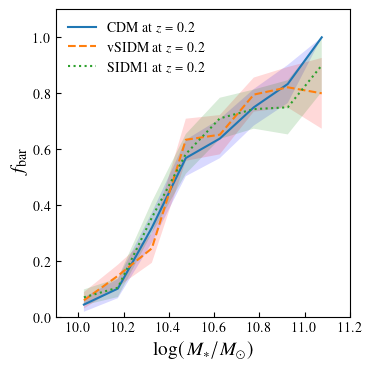}
  \caption{The bar fraction as a function of stellar mass, where stellar mass refers to the mass enclosed within a radius of 30 kpc. In each mass bin, following the expression $\Delta = \sqrt{{f_{\text{bar}}(1 - f_{\text{bar}})/n_{\text{disk}}}}$ is used to define the corresponding error region.}
 \label{fig1}
\end{figure}

  \begin{figure}
  \centering
    \includegraphics[width=0.45\textwidth]{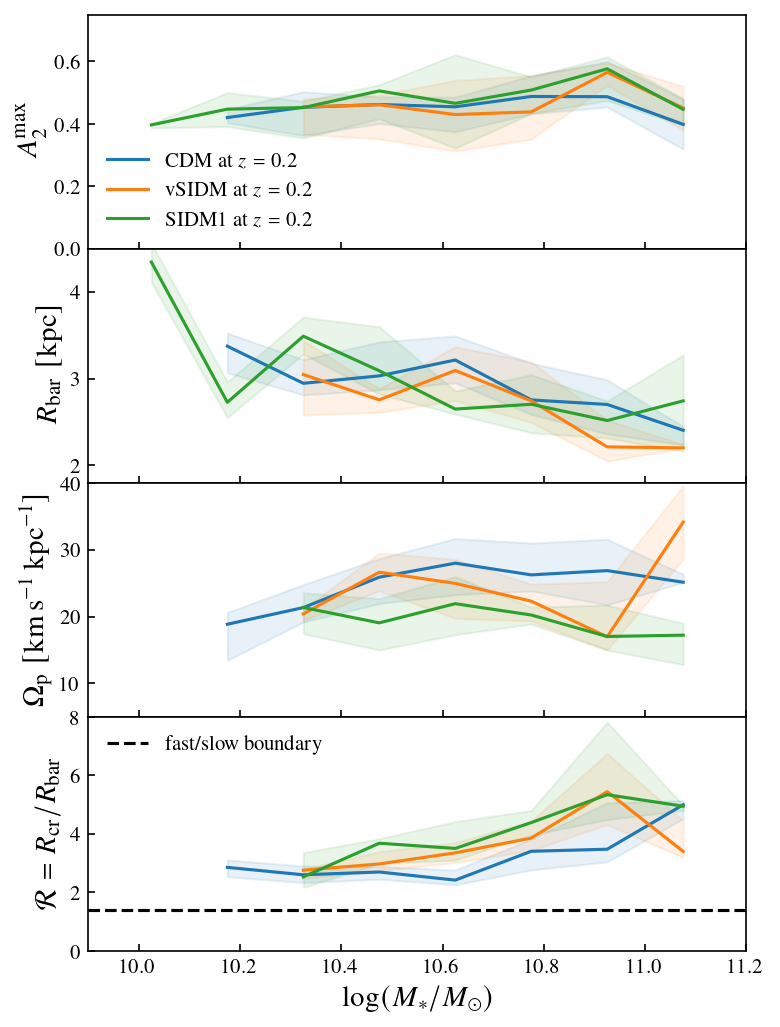}
  \caption{Top panel: bar strength, $A_2^{\rm max}$, as a function of stellar mass. The second panel shows the bar radius as a function of stellar mass. The third panel shows the bar pattern speed as a function of stellar mass. The fourth panel shows the speed parameter $\mathcal{R}$ as a function of stellar mass. All panels display data at redshift $z = 0.2$. The curves do not start from the same stellar mass because we omit mass bins containing fewer than three galaxies. The solid curves show the median values in each mass bin, and the shaded regions indicate the 32\textsuperscript{nd} to 68\textsuperscript{th} percentiles.}
 \label{fig2}
\end{figure}

\subsection{Pattern speed measurement}
\label{patt_s}
Although the Tremaine-Weinberg (TW) method \citep{tw1984} is widely used in both observations and simulations to measure bar pattern speeds (see \cite{roshan2025} and references therein), in this work we adopt the approach introduced in \cite{dehnen2024}. This newer method is fundamentally based on the same mathematical and physical intuition as the TW method, but it requires full six-dimensional phase-space information for the particles. Consequently, it cannot be applied to observational data for distant galaxies, where such complete information is unavailable. Therefore, its use is limited to simulations and possibly to the Milky Way and nearby galaxies, such as the Large Magellanic Cloud \citep{zhang2024,Arranz2024}. It is important to note that this method provides precise pattern speed measurements from a single snapshot \citep{dehnen2024,roshan2025}.

  \begin{figure*}
  \centering
    \includegraphics[width=0.33\textwidth]{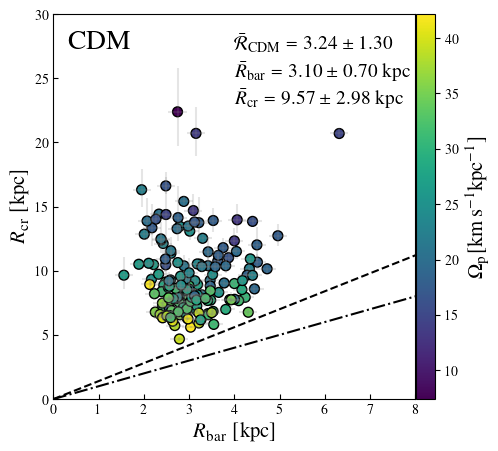}
    \includegraphics[width=0.33\textwidth]{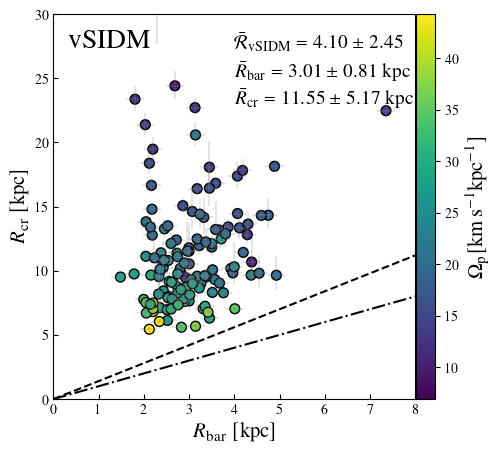}
    \includegraphics[width=0.33\textwidth]{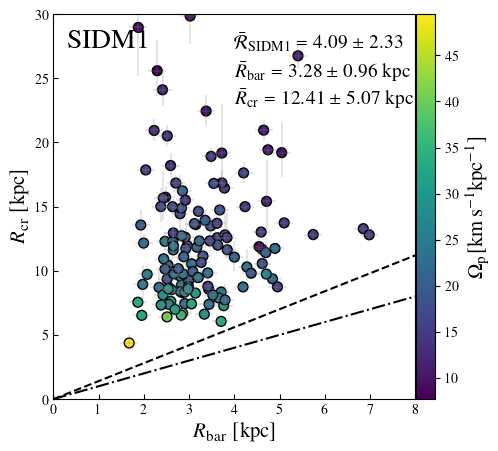}
  \caption{Distribution of the barred galaxies at $z=0.2$ in the $R_{\rm cr}-R_{\rm bar}$ plane. The dashed line corresponds to $\mathcal{R} = 1.4$, while the dot-dashed line corresponds to $\mathcal{R} = 1$. From left to right, we show CDM, vSIDM and SIDM1 galaxies. The colour bar shows the bar pattern speed $\Omega_{\rm bar}$.}
 \label{fig3}
\end{figure*}

\subsection{Corotation radius measurement}
For barred galaxies with a well-defined pattern speed $\Omega_{\rm p}$, we determine the corotation radius $R_{\rm cr}$ using the algebraic relation $\Omega_{\rm p} = \Omega(R_{\rm cr})$, where the angular velocity $\Omega(R)$ is obtained from the circular velocity of the galaxy $v_{\rm cir}(R)$. To obtain $v_{\rm cir}(R)$, we use the total mass enclosed within radius $R$, denoted as $M(R)$, which includes stellar, gas, and dark matter components. The circular velocity is then calculated as $v_{\rm circ}(R) = \sqrt{G M(R)/R}$. After measuring the corotation radius and the bar radius, we compute the parameter $\mathcal{R}$ for each galaxy. As mentioned earlier, this parameter is commonly used in observations to characterize the ``speed'' of the bars.

\section{Results}
\label{results}
Table \ref{tab:galaxy_counts} provides a concise summary of the relevant statistics. The second column, $N_{\rm gal}$, reports the number of galaxies with $M_* \geq 10^{10}\, M_\odot$ resolved with more than $10^4$ star particles identified in the CDM, vSIDM, and SIDM1 simulations. The third column, $N_{\rm disk}$, reports the number of disk galaxies passing the criteria defined in Sect. \ref{sample-selection}. This quantity exhibits a clear increasing trend with cosmic time, while showing similar values across different dark matter models at a given redshift. It is also interesting to note the behaviour of the disk fraction, $N_{\rm disk}/N_{\rm gal}$, which increases from approximately $30\%$ at $z=2$ to approximately $60\%$ at $z=0.2$ in all models. The fourth column, $N_{\rm bar}$, gives the number of barred galaxies, which likewise increases with time. The fifth column, $N_{\rm rel}$, lists the number of barred galaxies for which reliable measurements of pattern speed, bar length, and corotation radius are available. Pooled over all redshifts, a larger fraction of barred galaxies is excluded in going from $N_{\rm bar}$ to $N_{\rm rel}$ in the SIDM models than in CDM. In CDM, 49 of 376 barred galaxies (13.0\%) lack reliable measurements of the pattern speed, bar length, and corotation radius, compared with 66 of 382 (17.3\%) in SIDM1 and 75 of 382 (19.6\%) in vSIDM. This is mainly because the SIDM models contain more galaxies with very low pattern speeds and large measurement uncertainties. The final column, $N_{\rm matched}$, gives the number of matched galaxies in the simulations. The definition of matched galaxies will be given in Sect. \ref{sec:matched}. At each redshift, we begin by selecting galaxies that share the same initial conditions and then keep only those with reliable pattern speed, bar length, and corotation radius measurements in all three models. Matched galaxies that have reliable measurements in only one or two of the dark matter models are removed from the sample. In the following, we start by considering the results at the lowest redshift analyzed, $z=0.2$.

The bar fraction, defined as the ratio of barred galaxies to disk galaxies ($f_{\rm bar}=N_{\rm bar}/N_{\rm disk}$), is shown as a function of stellar mass in Figure~\ref{fig1}. Specifically, the bar fraction is calculated in mass bins of width $\Delta \log (M_*/M_{\odot}) = 0.15$, ensuring that each bin contains at least three galaxies. The stellar masses used are computed within a radius of 30 kpc. The binomial errors on the bar fractions are calculated using the bar fraction and the total number of disks, $n_{\text{disk}}$, in each mass bin, following the expression $\Delta = \sqrt{{f_{\text{bar}}(1 - f_{\text{bar}})/n_{\text{disk}}}}$ \citep{Rosas2022}. It is clear that, for all models, the bar fraction increases with stellar mass. This behavior has also been reported in the TNG50 simulation \citep{AHabibi}. However, the most important feature is that the bar fraction as a function of stellar mass is roughly similar across all models, with no significant differences observed among them. All disk galaxies with $A_2^{\rm max}\geq 0.2$ have been included in this plot.

The top panel of Figure~\ref{fig2} presents the bar strength, $A_2^{\rm max}$, as a function of stellar mass. This panel includes all $N_{\rm bar}$ disks with $A_2^{\rm max}\geq 0.2$, even though some may not have a well-defined pattern speed or bar length. The solid curves represent the median bar strength, and the shaded regions show the $32^{\text{nd}}$ to $68^{\text{th}}$ percentiles. Bars are generally expected to become stronger in more massive disks \citep{2016A&A...587A.160D}. Nevertheless, a decline in bar strength appears for $\log (M_*/M_{\odot}) \gtrsim 10.9$ across all models. The mean values of $A_2^{\rm max}$ for galaxies in Figure \ref{fig2} in the CDM, vSIDM, and SIDM1 models are $0.45 \pm 0.03$, $0.47 \pm 0.05$, and $0.47 \pm 0.05$, respectively, where the uncertainties denote the standard deviations.

   \begin{figure*}
  \centering
    \includegraphics[width=0.9\textwidth]{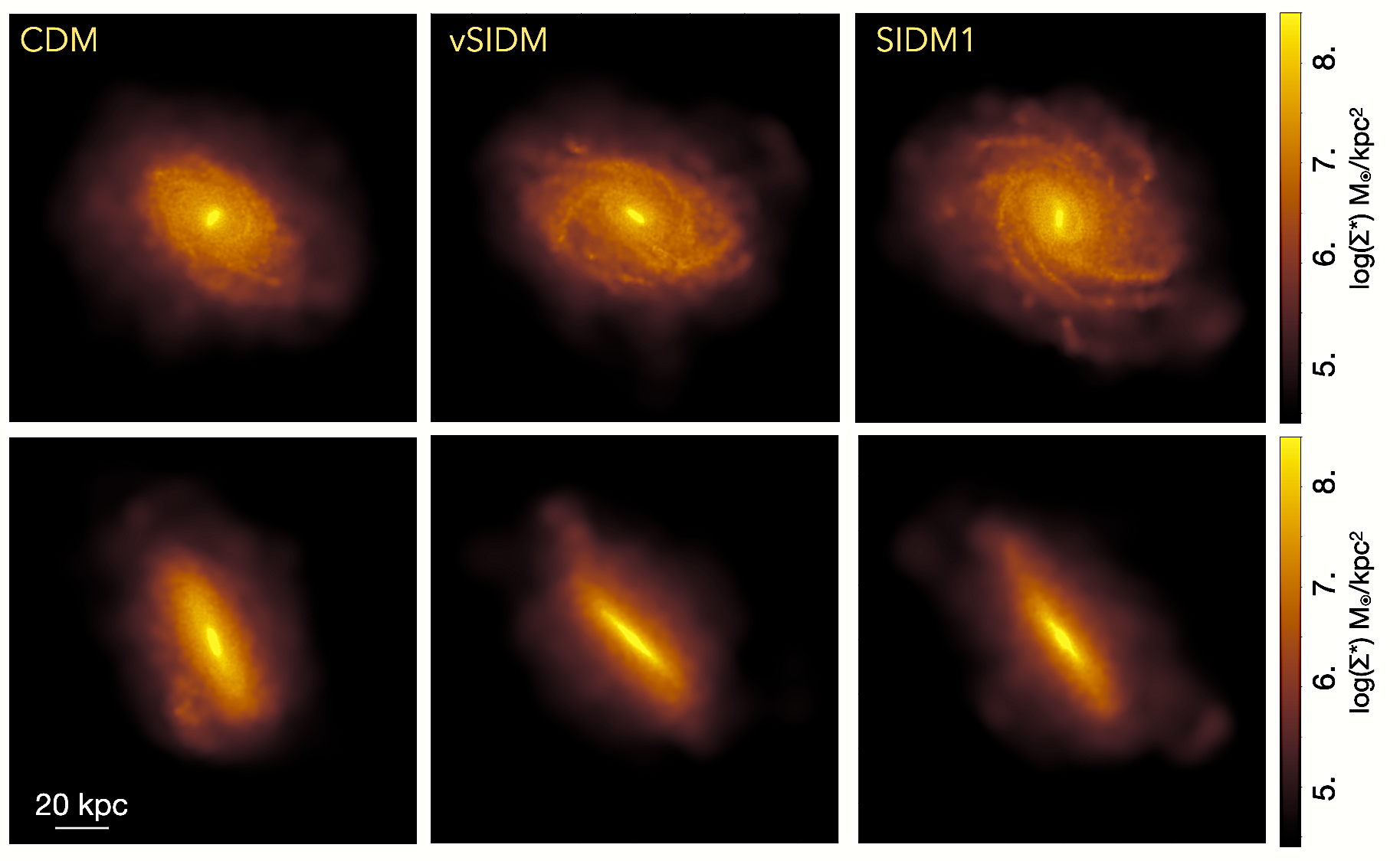}
   \caption{Visualization of the stellar projected density of a barred galaxy that matches across all three runs at $z=0.2$, viewed in the $xy$ and $xz$ projections which correspond roughly to face-on and edge-on configurations. The subhalo IDs are 76518, 75258 and 76450, respectively. The corresponding stellar masses are $\log(M_*/M_\odot)\simeq10.9$, $10.8$ and $10.8$; the bar radii are $R_{\rm bar}\simeq4.0$, $3.8$ and $3.7~{\rm kpc}$; and the pattern speeds are $\Omega_{\rm bar}\simeq31.0$, $20.5$ and $17.8~{\rm km\,s^{-1}\,kpc^{-1}}$.}
 \label{fig_vis1}
\end{figure*}

The second panel of Figure~\ref{fig2} shows the bar radius, defined as half of the bar length. The mean values of $R^{\rm bar}$ in the CDM, vSIDM, and SIDM1 models are $2.92 \pm 0.33$, $2.68 \pm 0.39$, and $3.03 \pm 0.61$ kpc, respectively. The bar radius clearly decreases with increasing stellar mass independently of the dark matter model. Vice versa, observations of nearby galaxies reveal a strong positive correlation between bar length and stellar mass, with bar length increasing with stellar mass (see Figure 3 in \cite{AHabibi} where the dashed curve belongs to observation). Figure~\ref{fig2} includes all barred galaxies with reliable bar-length measurements. Specifically, the CDM, vSIDM, and SIDM1 models contain 165, 159, and 149 such galaxies, respectively.  A comparable problem is found in TNG50, where bar length varies only weakly with stellar mass; see Figure~3 of \cite{AHabibi} for the bar radius as a function of stellar mass at three redshifts. IllustrisTNG has difficulty reproducing observed bar sizes. \citet{2022ApJ...940...61F} likewise reported this for TNG50, comparing the bar-size distribution at $z = 0$ with that of MaNGA observations. They found that TNG50 bars are on average about 35\% shorter than observed. This finding suggests that the fast-bar tension may be related primarily to the bar size (the scarcity of long bars in TNG50) rather than to the pattern speed itself.

The third panel of Figure~\ref{fig2} shows the pattern speeds of bars in the CDM, vSIDM, and SIDM1 models as a function of stellar mass. The mean values of $\Omega^{\rm bar}$ in the CDM, vSIDM, and SIDM1 models are $24.60 \pm 3.29$, $24.22 \pm 5.92$, and $19.46 \pm 2.07\, {\rm km\, s^{-1}\, kpc^{-1}}$, respectively. The pattern speeds in the vSIDM and SIDM1 models are clearly lower than those in the CDM model for stellar masses below $10^{10.9}\, M_{\odot}$, with the difference being more pronounced in SIDM1. At masses above $10^{10.9}\, M_{\odot}$, the pattern speeds in vSIDM become higher than those in CDM. This panel includes all $N_{\rm rel}$ barred galaxies with reliable pattern speed, bar length and corotation radius measurements. Galaxies with relative errors exceeding 20\% are excluded. The numbers of galaxies excluded because of this limit on pattern speed are 9 for CDM, 20 for vSIDM, and 21 for SIDM1\footnote{Applying this criterion to other redshifts, the numbers of galaxies excluded from the CDM sample at $z = 1$, $1.5$, and $2$ are $(10, 5, 2)$, respectively. The corresponding numbers are $(13, 4, 8)$ for vSIDM and $(9, 7, 0)$ for SIDM1. With the exception of vSIDM at $z = 2$, the fraction of excluded galaxies is below $14\%$. This exclusion is unlikely to introduce a model-dependent selection bias, since it removes only those galaxies whose pattern speeds cannot be measured reliably, and the pattern speeds are measured with the same method in all models.}. 

The bottom panel presents the $\mathcal{R}$ parameter. The mean values of $\mathcal{R}$ in the CDM, vSIDM, and SIDM1 models are $3.21 \pm 0.88$, $3.63 \pm 0.97$, and $4.07 \pm 1.03$, respectively. Although all models remain in the slow regime, the SIDM models show even slower bar rotation, at least for masses below $10^{10.9}\, M_{\odot}$. The horizontal dashed line indicates the division between the fast and slow regimes, i.e. $\mathcal{R}=1.4$. All the galaxies with reliable pattern speed, corotation and bar length measurement have been included in this panel.

The distribution of galaxies in the $R_{\rm cr}$-$R_{\rm bar}$ plane is shown in Figure~\ref{fig3} for all $N_{\rm rel}$ galaxies with reliable measurements of pattern speed, bar length, and corotation radius. The dashed line corresponds to $\mathcal{R} = 1.4$, which separates the fast and slow bar regions, while the dot-dashed line corresponds to $\mathcal{R} = 1$, indicating the ultra-fast bar region. The color bars represent the pattern speed of the bars. As expected from Figure \ref{fig2}, galaxies in the CDM model tend to have smaller corotation radii. The mean value of the bar radius $R_{\rm bar}$, corotation radius $R_{\rm cr}$ and $\mathcal{R}$ parameter, together with the corresponding standard deviations, are displayed in the upper right corner of each panel. We compared the distributions of the bar corotation parameter $\mathcal{R}$ among the vSIDM, SIDM1, and CDM models using pairwise two-sample Kolmogorov-Smirnov (KS) tests. The vSIDM and SIDM1 distributions are statistically consistent with each other ($p=0.59$), whereas both differ significantly from the CDM distribution ($p=1.8\times10^{-5}$ for vSIDM versus CDM and $p=2.8\times10^{-5}$ for SIDM1 versus CDM). Adopting a significance threshold of $p<0.05$, we conclude that the $\mathcal{R}$ distributions of the two SIDM models are mutually consistent but differ significantly from that of CDM.
\begin{figure*}
\centering
    \includegraphics[width=0.227\textwidth]{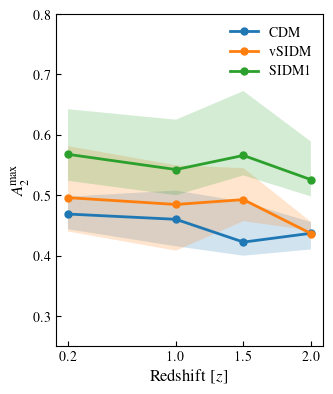}
    \includegraphics[width=0.233\textwidth]{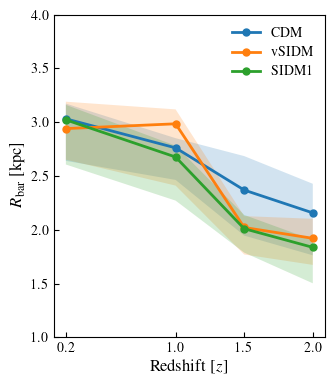}
    \includegraphics[width=0.233\textwidth]{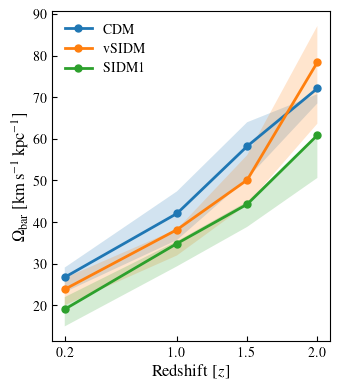}
    \includegraphics[width=0.233\textwidth]{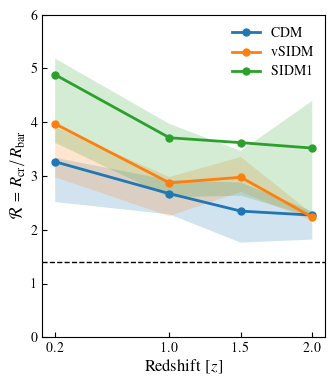}
 \caption{From left to right, the first row shows the bar strength, bar radius, bar pattern speed, and the dimensionless bar-speed parameter $\mathcal{R}$ for the matched galaxy sample as a function of redshift in the three dark matter models considered.}
 \label{fig4p}
\end{figure*}

\begin{figure*}
\centering
    \includegraphics[width=0.24\textwidth]{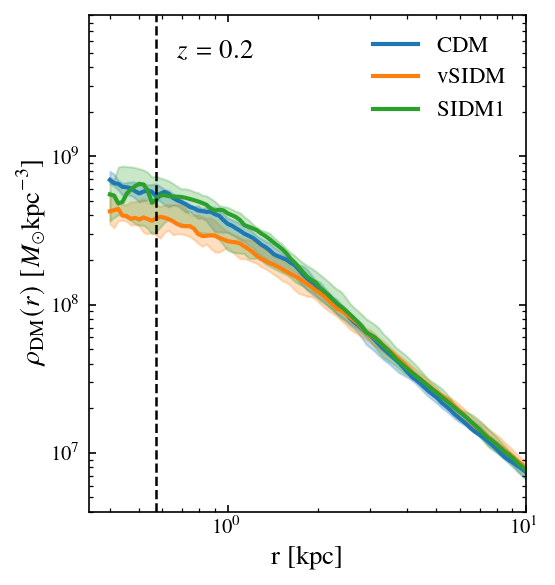}
    \includegraphics[width=0.24\textwidth]{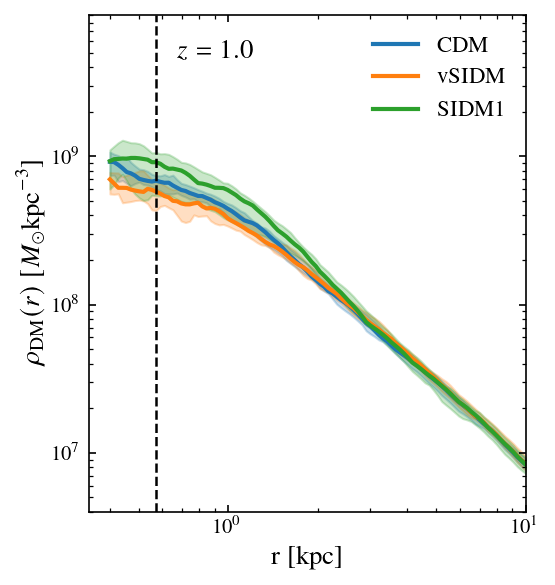}
    \includegraphics[width=0.24\textwidth]{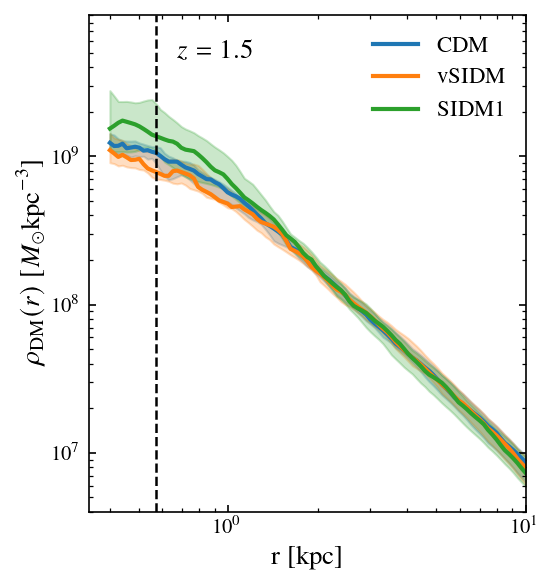}
    \includegraphics[width=0.24\textwidth]{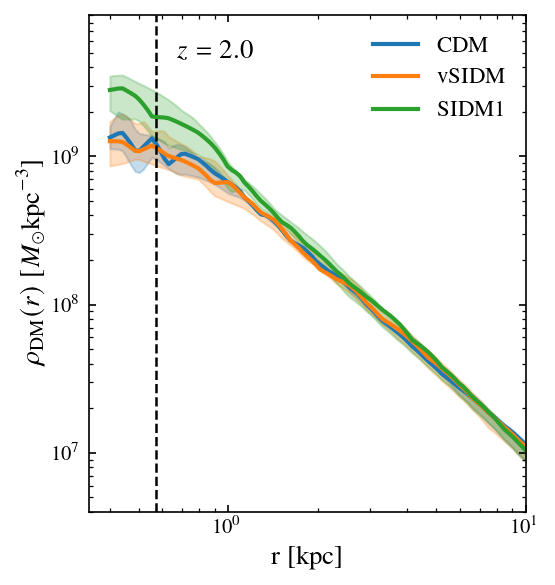}\\
    \includegraphics[width=0.24\textwidth]{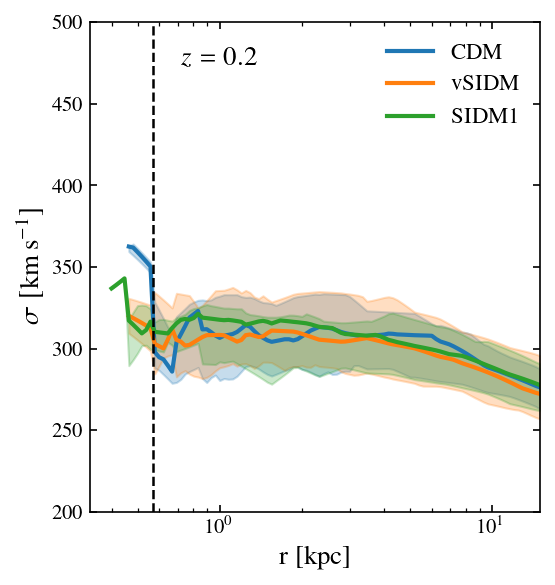}
    \includegraphics[width=0.24\textwidth]{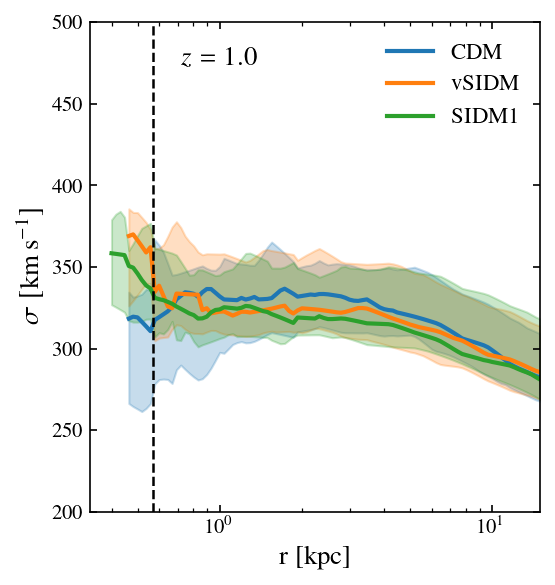}
    \includegraphics[width=0.24\textwidth]{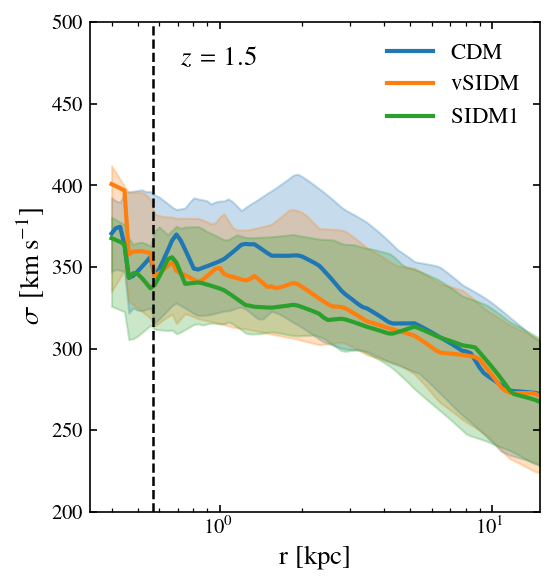}
    \includegraphics[width=0.24\textwidth]{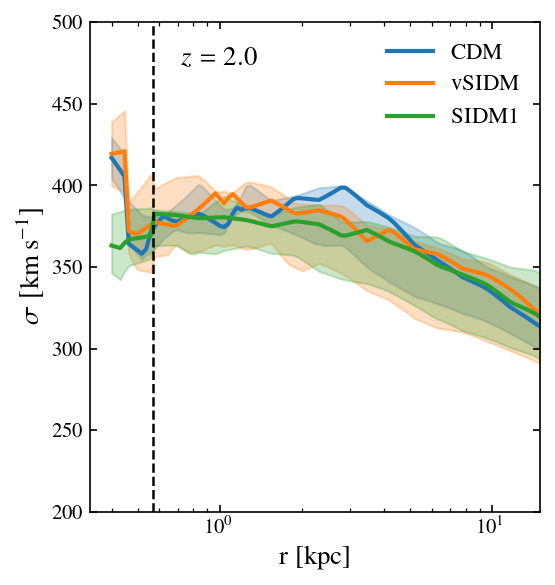}\\
    \includegraphics[width=0.24\textwidth]{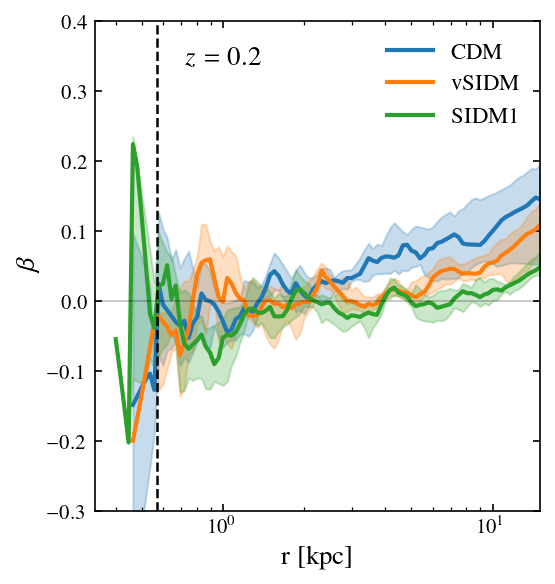}
    \includegraphics[width=0.24\textwidth]{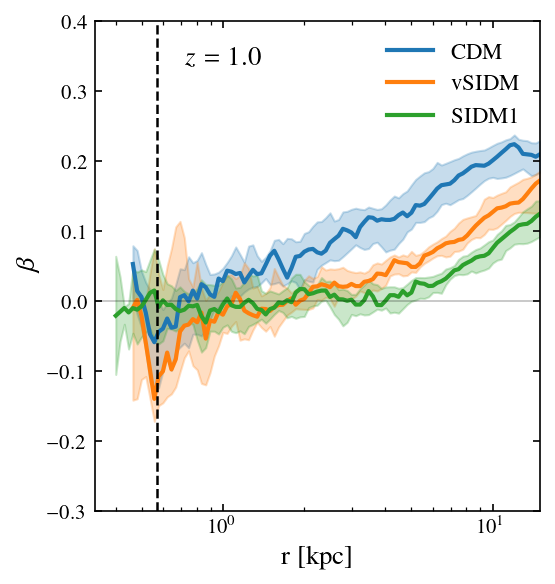}
    \includegraphics[width=0.24\textwidth]{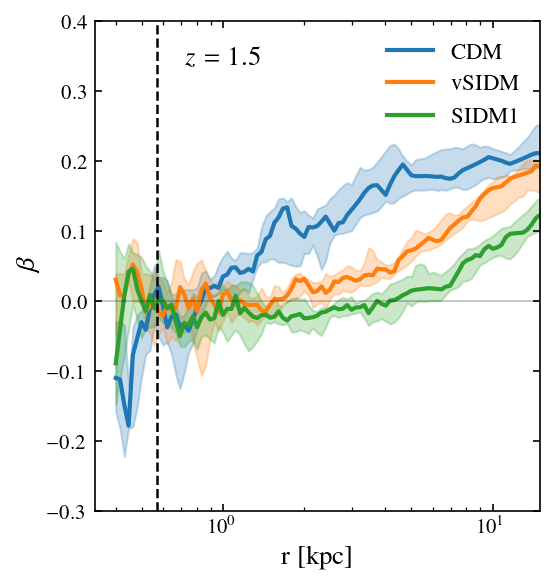}
    \includegraphics[width=0.24\textwidth]{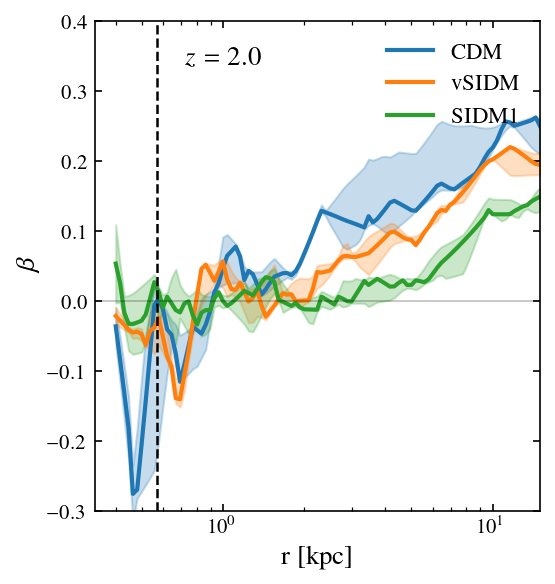}
 \caption{The second row presents the mean dark matter density profiles at several redshifts. The third row shows the corresponding total dark matter velocity-dispersion profiles, $\sigma$, while the fourth row displays the velocity-anisotropy parameter,
$\beta=1-(\sigma_{\theta}^2+\sigma_{\phi}^2)/(2\sigma_r^2)$, evaluated in spherical coordinates. In all radial-profile panels, the vertical dashed line marks $r=\epsilon_{\rm DM}$, where $\epsilon_{\rm DM}=0.57\,{\rm kpc}$ is the gravitational softening
length of the AIDA--TNG simulations \citep{despali2025}.}
 \label{fig4}
\end{figure*}

\subsection{Matched galaxies}
\label{sec:matched}
To enable a more meaningful comparison among the dark matter models considered in this study, we focus on galaxies that originate from identical initial conditions but evolve under different dark matter prescriptions. We refer to these as matched galaxies. Every galaxy in a given dark matter model has a counterpart in each of the other two models; however, these counterparts do not necessarily share the same morphology. We therefore retain only those systems that are classified as barred disks in all three models. 
At each redshift, we identify galaxies that share the same initial conditions and retain only those with reliable measurements of pattern speed, bar length, and corotation radius in all three models. Matched galaxies with reliable measurements in only one or two of the dark matter models are removed from the sample. The number of matched galaxies is listed in the last column of Table \ref{tab:galaxy_counts}. In Figure \ref{fig_vis1}, we show an example of matched barred galaxies, highlighting the stellar distribution where the central bar is clearly visible.

Figure~\ref{fig4p} presents, from left to right, the mean bar strength, bar radius, bar pattern speed, and speed parameter $\mathcal{R}$ as a function of redshift for the matched galaxy sample. The first panel
clearly shows that bars are strongest in the SIDM1 model, with bars in vSIDM also exhibiting greater strength than their CDM counterparts. Combining all redshifts, we compared the distributions of $A_2^{\rm max}$ among the models using KS test. The CDM and vSIDM distributions are statistically consistent with each other ($p \simeq 0.05$), whereas CDM differs significantly from SIDM1 ($p = 7.4\times10^{-9}$). The two SIDM models also differ significantly from each other ($p = 3.5\times10^{-4}$). Thus, at a significance threshold of $p < 0.05$, SIDM1 produces a distinctly different $A_2^{\rm max}$ distribution, while vSIDM remains statistically consistent with CDM.

 The second panel reveals no monotonic difference in bar radius among the three models, though bars tend to be longer in CDM at high redshifts ($z = 1.5$ and $z = 2$). The KS test shows that the bar radii in these models are statistically consistent with one another. The third panel shows the mean pattern speed; bars in CDM rotate faster at all redshifts, with the exception of $z = 2$ where the sample is limited to only 5 matched galaxies, while bars in SIDM1 are the slowest. However, combining all redshifts, the KS test reveals that the CDM and vSIDM distributions are statistically consistent with each other ($p = 0.36$). In contrast, CDM differs significantly from SIDM1 ($p = 8.9\times10^{-3}$) as are the vSIDM and SIDM1 distributions ($p = 3.5\times 10^{-4}$). Thus, at a significance threshold of $p < 0.05$, the bar pattern-speed distribution of SIDM1 differs from that of CDM, while vSIDM remains statistically consistent with CDM.

This hierarchy is consistently reflected in the fourth panel in terms of the speed parameter $\mathcal{R}$. Overall, bars in SIDM models rotate more slowly, both in terms of pattern speed magnitude and the value of $\mathcal{R}$, with the ordering from slowest to fastest being SIDM1, vSIDM, and CDM. However, combining all redshifts, the KS test reveals that the CDM and vSIDM distributions are statistically consistent with each other ($p=0.11$). In contrast, CDM differs significantly from SIDM1 ($p=1.3\times10^{-6}$), and vSIDM also differs significantly from SIDM1 ($p=3.3\times10^{-3}$). Thus, at a significance threshold of $p<0.05$, SIDM1 produces a distinctly different $\mathcal{R}$ distribution from both CDM and vSIDM, while vSIDM remains statistically consistent with CDM. Therefore, the fast-bar tension identified in CDM \citep{2021MNRAS.508..926R} appears to be even more pronounced in SIDM models specifically in SIDM1.

\subsubsection{Dynamical friction}
In all three models, the bar pattern speed declines monotonically as a function of cosmic time, with dynamical friction from the dark matter halo serving as a principal driver of this deceleration \citep{AHabibi}. We quantify the overall evolution through the fractional change $\Delta = (\Omega_i - \Omega_f)/\Omega_i$, where $i$ and $f$ denote the initial and final epochs, respectively. Evaluating this quantity between $z=2$ and $z=0.2$, we obtain $\Delta_{\rm CDM}\simeq 61\%$, $\Delta_{\rm vSIDM}\simeq 68\%$, and $\Delta_{\rm SIDM1}\simeq 69\%$. These results suggest that the mean rate of decline in bar angular velocity is roughly
$10\%$ larger in the SIDM models, implying that the exchange of angular momentum between the bar and the halo operates more efficiently in SIDM than in CDM. This is consistent with the results of \cite{bosch2026}, who find that angular momentum exchange is enhanced in SIDM. Employing a suite of high-resolution idealized $N$-body simulations complemented by an analytical framework grounded in kinetic theory, they demonstrated that, relative to collisionless CDM halos, SIDM halos systematically promote earlier bar formation and produce stronger bars, even for modest self-interaction cross sections. This finding is consistent with our results from AIDA-TNG, where bars are likewise found to be stronger in SIDM models. They further showed that dark matter self-interactions broaden the bar--halo resonances, enabling a larger fraction of halo particles to participate in resonant angular momentum exchange, thereby enhancing the efficiency of angular momentum transfer from the stellar disk to the dark matter halo.

Since only five matched galaxies exist at $z=2$, we additionally evaluate $\Delta$ using $z=1.5$ as the initial redshift, yielding $\Delta_{\rm CDM}\simeq 52\%$, $\Delta_{\rm vSIDM}\simeq 51\%$, and $\Delta_{\rm SIDM1}\simeq 58\%$. Here, no appreciable difference emerges between CDM and vSIDM, while the decline rate in SIDM1 remains $\sim$ 10\% higher. Across both redshift baselines, the results are broadly consistent among all models, suggesting that the mean rate of decline in bar angular velocity is not significantly model-dependent and that the integrated efficiency of dynamical friction is comparable across the dark matter scenarios considered. This consistency does not imply, however, that the functional form of dynamical friction is the same in each model. Dynamical friction is shaped by the complex interplay of several physical quantities, including the dark matter density, the bar strength, the velocity dispersion of dark matter particles, encapsulated, for the point-mass perturber case, within the Chandrasekhar formula, and the degree of anisotropy in the velocity distribution \citep{binney1977,kroupa2004}. As we discuss further in this section, these quantities differ in physically meaningful ways between CDM and the SIDM variants considered here.
\begin{figure*}
\centering
    \includegraphics[width=0.227\textwidth]{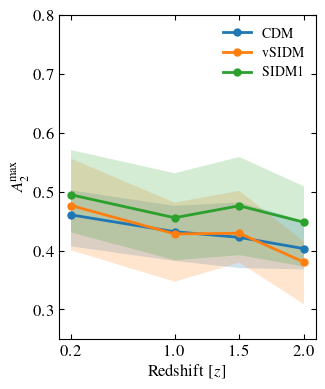}
    \includegraphics[width=0.237\textwidth]{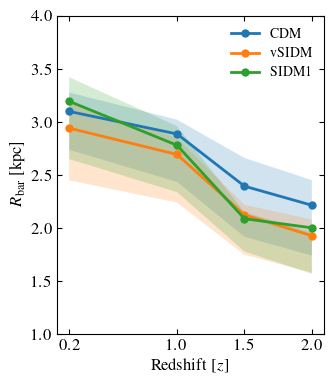}
    \includegraphics[width=0.233\textwidth]{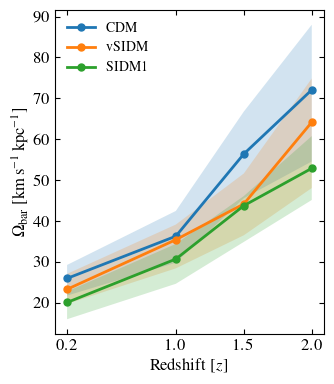}
    \includegraphics[width=0.237\textwidth]{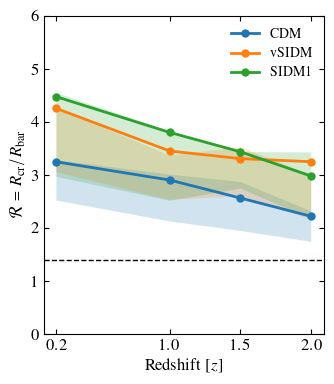}
 \caption{Similar to Figure~\ref{fig4p}, but here we include all $N_{\rm rel}$ barred galaxies with reliable pattern speed and bar length measurements, without restricting the sample to matched galaxies.}
 \label{fig5p}
\end{figure*}

\begin{figure*}
\centering
    \includegraphics[width=0.24\textwidth]{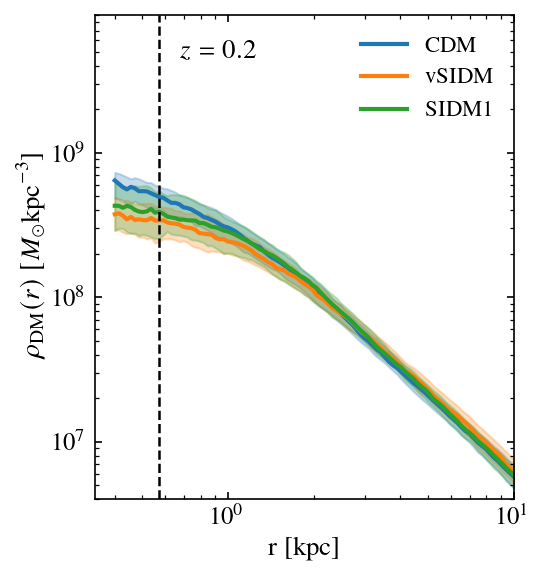}
    \includegraphics[width=0.24\textwidth]{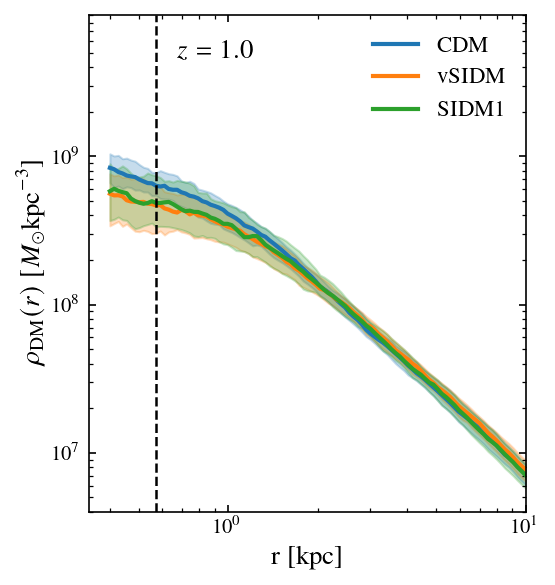}
    \includegraphics[width=0.24\textwidth]{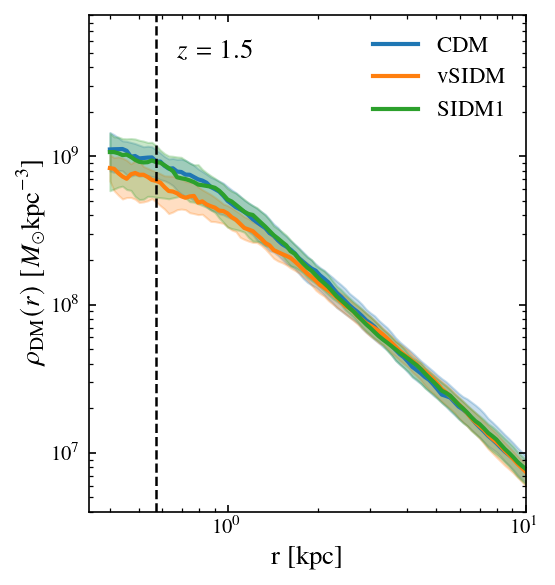}
    \includegraphics[width=0.24\textwidth]{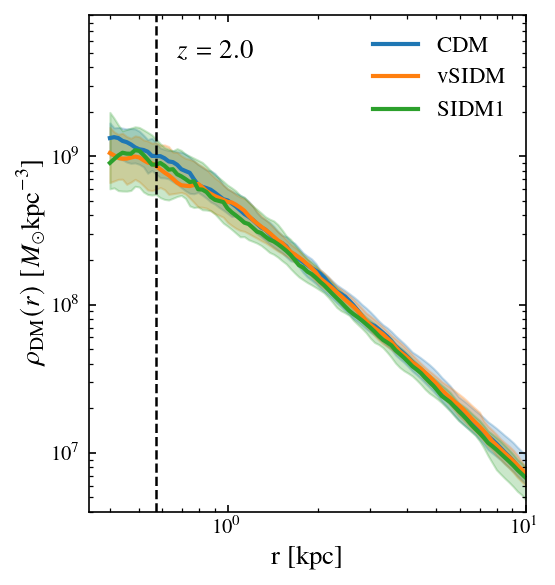}\\
    \includegraphics[width=0.24\textwidth]{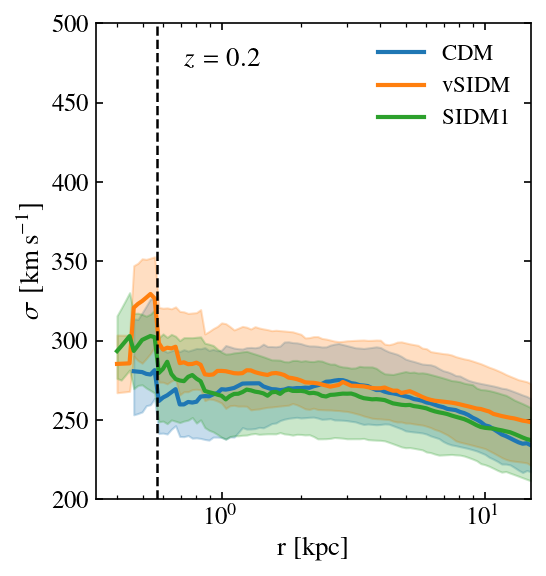}
    \includegraphics[width=0.24\textwidth]{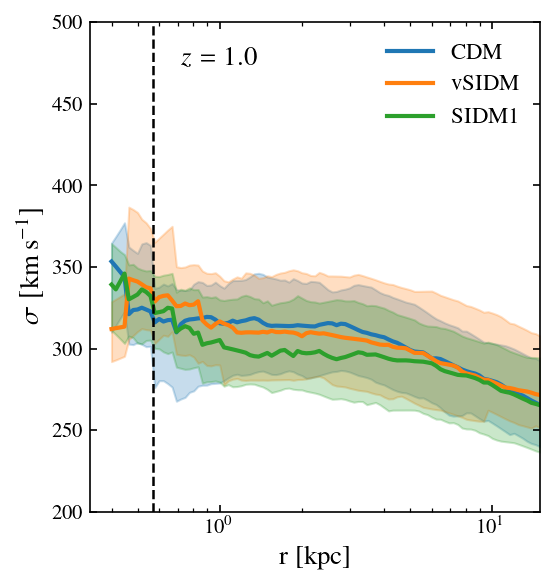}
    \includegraphics[width=0.24\textwidth]{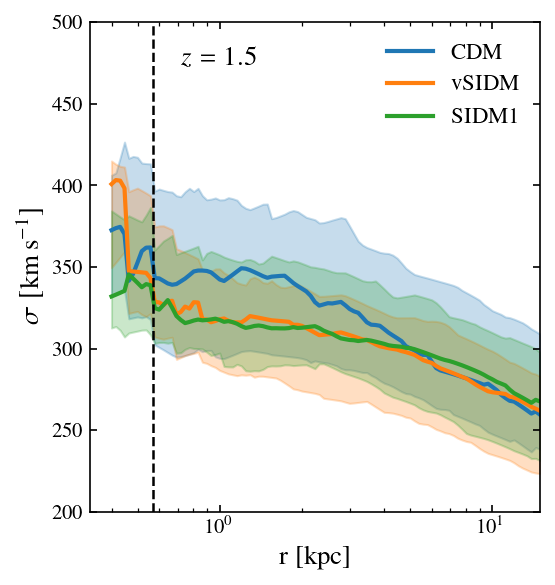}
    \includegraphics[width=0.24\textwidth]{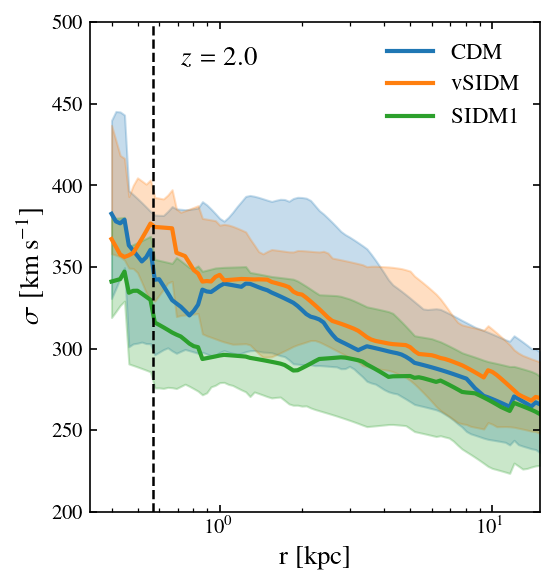}\\
    \includegraphics[width=0.24\textwidth]{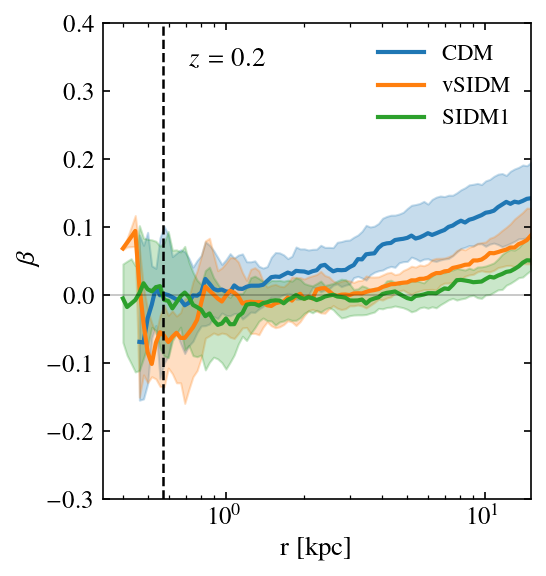}
    \includegraphics[width=0.24\textwidth]{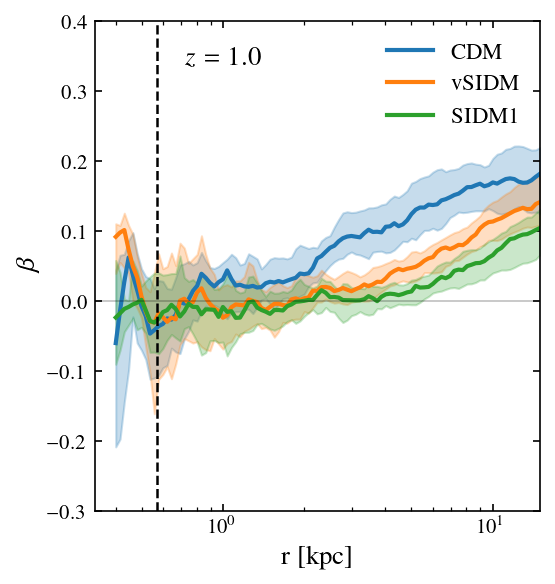}
    \includegraphics[width=0.24\textwidth]{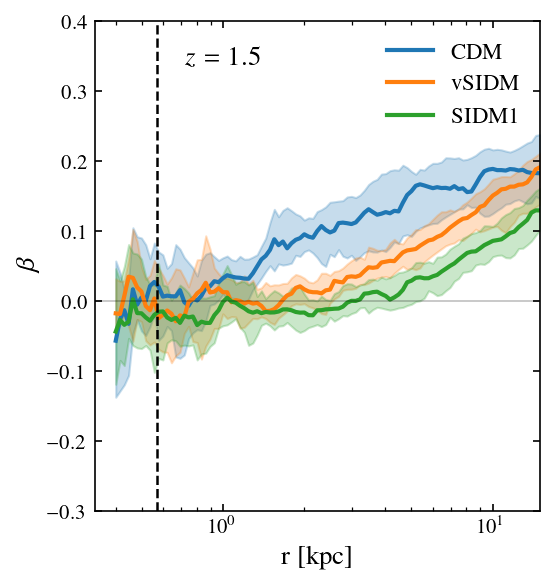}
    \includegraphics[width=0.24\textwidth]{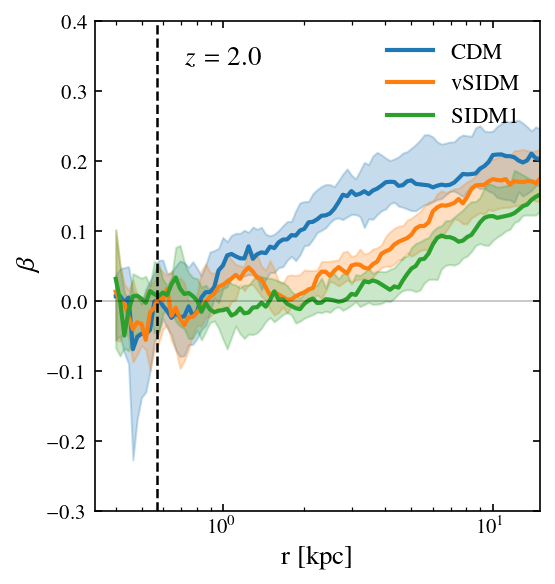}
 \caption{Similar to Figure~\ref{fig4}, but here we include all $N_{\rm rel}$ barred galaxies with reliable pattern speed and bar length measurements, without restricting the sample to matched galaxies.}
 \label{fig5}
\end{figure*}

The first row of Figure~\ref{fig4} presents the mean dark matter halo density profiles at various redshifts for the matched galaxies. At radii smaller than $\epsilon_{\rm DM}$, the measured halo profiles may be affected by gravitational softening, numerical
two-body relaxation, and resolution-dependent artificial flattening. We therefore restrict the physical interpretation and quantitative comparison of the profiles to $r>\epsilon_{\rm DM}$. The profiles at smaller radii are shown only for completeness. This choice is consistent with the radial range adopted by \cite{despali2026} for fitting the AIDA-TNG halo profiles. At $r>3\epsilon_{\rm DM}$, no statistically significant difference is found among the mean dark matter density profiles of the three models. However, in the radial interval $\epsilon_{\rm DM} < r < 3\epsilon_{\rm DM}$, we have $\rho_{\rm SIDM1} > \rho_{\rm CDM} > \rho_{\rm vSIDM}$ at all redshift. Nevertheless, the KS test reveals that the density profiles across all models remain statistically consistent across nearly all redshifts\footnote{To quantify differences in the density profiles among the dark matter models, we reduced the density profile of each galaxy to a single number, namely its median value in the radial range $0.57<r<10$~kpc. We then applied a two-sample KS test to each pair of models, comparing the distributions of these per-galaxy medians. Because the test acts on one summary statistic per galaxy, it compares the distributions of galaxies across models rather than the shapes of individual profiles. A similar procedure is taken for the dispersion velocity and the velocity anisotropy. The qualitative results are insensitive to the reasonable choices of the radial range adopted above.}. The sole exception occurs at $z = 1.0$, where SIDM1 deviates significantly from both CDM and vSIDM. Consequently, halo density variations cannot be invoked as the primary driver for the differing dynamical friction among these models. However, \citet{despali2026} also reported that a subset of SIDM1 galaxies shows steeper inner profiles compared to CDM and that this effect correlates with a smaller stellar effective radius; it is thus interesting that these appear to be the galaxies that belong to the matched sample, even if the origin of the effects still needs to be identified.

The second row of Figure~\ref{fig4} shows the mean total velocity dispersion, $\sigma$, of the halo particles at various redshifts. As discussed above, the velocity dispersion directly modulates the magnitude of dynamical friction. As the standard Chandrasekhar expression indicates, increasing $\sigma$ suppresses the magnitude of dynamical friction. Although the Chandrasekhar formula was derived for a point-mass perturber, \cite{sellwood2006} demonstrated that the frictional torque on a rotating, imposed bar scales with halo properties in a manner well described by the Chandrasekhar framework. The KS test demonstrates that the velocity dispersion profiles remain statistically consistent across all models and redshifts. Consequently, analogous to our findings for the dark matter density, dark matter velocity dispersion cannot be considered as a key driver of the differing dynamical friction among the models.

The third row of Figure~\ref{fig4} presents the mean anisotropy parameter, $\beta=1-(\sigma_{\theta}^2+\sigma_{\phi}^2)/2\sigma_r^2$, evaluated in spherical coordinates. Positive values of $\beta$ across all models confirm that the dark matter halos are radially anisotropic. As discussed above, velocity anisotropy can affect dynamical friction through both direct and indirect channels. \cite{sellwood2016} showed that, in the CDM framework, radially biased halos give rise to bars with reduced growth rates. The fourth row reveals a clear ordering: $\beta$ is highest in CDM, intermediate in vSIDM, and lowest in SIDM1. All profiles are radially biased ($\beta > 0$), at least in the outer regions of the bar at $r \gtrsim 5\,{\rm kpc}$. Clearly, in the inner regions the SIDM1 halos are more isotropic ($\beta \simeq 0$) than those of the other two models. More importantly, pairwise KS tests demonstrate statistically significant differences in the $\beta$ profiles across all models at every redshift. In accordance with \cite{sellwood2016}, this ordering implies that bars in SIDM models are expected to be stronger than their CDM counterparts, consistent with the bar strength results, at least for the SIDM1 case, presented in the left panel of the first row. Stronger bars naturally couple more efficiently to the halo, thereby experiencing larger frictional torques, this constitutes the indirect influence of anisotropy on dynamical friction. Beyond this indirect effect, the degree of velocity anisotropy is also known to directly affect the magnitude of dynamical friction \citep{binney1977}, so the systematic differences in $\beta$ seen across models may have an additional direct bearing on the dynamical friction efficiency. In a recent study, \citet{Gherghinescu2026} employed controlled simulations in which the galactic and bar potentials are imposed, while test-particle dynamics are followed only for the halo particles. They showed that the velocity anisotropy plays a crucial role in setting the strength of the resonant torques on the halo, as described by the Lynden-Bell--Kalnajs torque formalism \citep{1972MNRAS.157....1L}. More specifically, the more radially biased the halo, the stronger the retrograde torque at the 1:1 resonance. At this resonance, the halo loses angular momentum owing to a larger imbalance in the gradients of the phase-space distribution function across the resonant transport direction in action space. Consequently, the net torque exerted by the halo on the bar, and hence the rate of bar slowdown, cannot be inferred from the halo density profile alone. Indeed, two halos with identical density distributions but different velocity-anisotropy structures can couple to the bar with substantially different torque strengths, since the relevant quantity is the gradient of the distribution function across the resonant directions in action space rather than the bulk velocity dispersion. This dependence is directly relevant to comparisons between CDM and SIDM halo models: differences in the predicted anisotropy profiles could translate into different bar--halo torque histories, even when the density profiles are matched.

Taken together, the interplay of all the key physical factors discussed above yields an angular velocity decline rate in SIDM1 that is approximately 10\% higher than in the other two models. We emphasize that our analysis showed that the dark matter density, velocity dispersion, and bar size are statistically consistent among the CDM and SIDM models. In contrast, the velocity anisotropy differs markedly between the models, which makes it a possible explanation for the differences in dynamical friction among them. As discussed above, velocity anisotropy influences the angular momentum transfer both directly, by modifying the strength of the bar-halo resonant torques, and indirectly, by affecting bar formation and the resulting bar strength. Bars are stronger in the SIDM1 model, and, as noted earlier, dynamical friction also appears to operate more efficiently in this model.

It is important to note, however, that the systematically lower values of $\Omega_{\rm p}$ and accordingly higher values of $\mathcal{R}$ found in SIDM models at low redshift ($z=0.2$) cannot be ascribed exclusively to dynamical friction. In SIDM models, the pattern speeds are already lower at the earliest epochs at which bars are identified in our sample, and this offset persists to lower redshifts. Whether this reflects a lower pattern speed at the onset of bar formation cannot be established with the present data, since we do not identify the epoch of bar formation for individual systems. The more fundamental question is therefore why the pattern speeds are already lower at these early epochs in SIDM models.
The answer is likely encoded in the dark matter halo properties discussed above, though a definitive assessment would require a complementary study covering a broader redshift baseline extending beyond $z=2.0$, identifying the epoch of bar formation across a statistical sample of barred galaxies, and comparing the baryonic and dark matter properties at that epoch across different dark matter models. We reserve this for a future study. The central result of this paper, however, is that the fast-bar tension previously reported in standard cold dark matter cosmological simulations, see \cite{2021MNRAS.508..926R,Fragkoudi2021,fire2}, persists in SIDM models, and is even more pronounced.

As a final remark in this section, since the number of matched galaxies is relatively small, the results presented for the matched sample may be affected by small-number statistics. We show that this is not the case and that our main results remain qualitatively unchanged when a larger sample is considered. To this end, Figures~\ref{fig5p} and \ref{fig5} are analogous to Figures~\ref{fig4p} and \ref{fig4}, but include all $N_{\rm rel}$ barred galaxies at the different redshifts. For this larger sample of barred galaxies, the KS test results for $A_2^{\rm max}$ are the same as those obtained for the matched sample shown in Figure~\ref{fig4p}. For $\Omega_{\rm p}$, however, the distributions in the vSIDM and SIDM1 models are statistically
consistent, unlike those in the matched sample. Also in contrast to the matched sample, the $R_{\rm bar}$ distribution in CDM differs significantly from those in the vSIDM and SIDM1 models, with CDM bars being a bit longer. The distribution of $\mathcal{R}$ in CDM also differs significantly from those in vSIDM and SIDM1, with bars in the SIDM models exhibiting larger values of $\mathcal{R}$. For the dark matter density, and unlike in the matched sample, the CDM and vSIDM distributions are statistically different at all redshifts except $z=2.0$, with lower inner halo densities in vSIDM. As for the matched galaxies, the halo velocity dispersion $\sigma$ is statistically consistent among all models at all redshifts, except for the CDM and vSIDM pair at $z=0.2$. Finally, and more importantly, as in the matched sample, the anisotropy parameter $\beta$ is statistically different among all models at all redshifts.

\section{Conclusion}
\label{conclusion}
We have presented the first systematic measurement of galactic bar pattern speeds in SIDM cosmological simulations. Using the $51.7\,{\rm Mpc}$ box of the AIDA-TNG project \citep{despali2025}, we identified barred disk galaxies with $M_* > 10^{10}\,M_{\odot}$ in the CDM, SIDM1, and vSIDM runs at redshifts $z=0.2,\,1.0,\,1.5$, and $2.0$. We compared their statistical and dynamical bar properties across the three dark matter models, both at the population level and for galaxies matched across the runs, taking advantage of their identical initial conditions. The bar fraction as a function of stellar mass is essentially indistinguishable among the three models, indicating that dark matter self interactions do not strongly affect the overall incidence of bars in this mass range. In other words, the population statistics of barred galaxies are similar in all models, while the bar dynamics differ systematically.
  \begin{figure}
  \centering
  \includegraphics[width=0.4\textwidth]{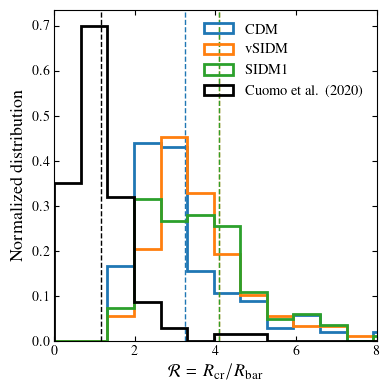}
  \caption{Distribution of the $\mathcal{R}$ parameter at $z=0.2$ for the CDM, vSIDM, and SIDM1 models. The black histogram shows the observed distribution from \citet{Cuomo2020} for nearby galaxies. The vertical dashed lines mark the mean value of each distribution. }
 \label{fig7}
\end{figure}
 
The bars in the SIDM models rotate more slowly than their CDM counterparts. More specifically, when all redshifts are combined, the pattern speed distribution in the CDM model differs significantly from that in SIDM1, with bars in SIDM1 rotating more slowly. Correspondingly, KS tests at $z=0.2$ show that the $\mathcal{R}$ parameter differs significantly between the CDM and SIDM models, with systematically higher values in the SIDM models. The distributions of this parameter are shown in Figure~\ref{fig7}, where the vertical dashed lines mark the mean value of each distribution. The fast-bar tension reported in standard cold dark matter cosmological simulations \citep{2021MNRAS.508..926R,Fragkoudi2021,fire2} therefore persists in SIDM models and is further aggravated. We examined whether these differences can be attributed to dark matter dynamical friction. The fractional decline of the pattern speed between $z=2$ and $z=0.2$ is modestly ($\sim 10\%$) larger in SIDM1 than in CDM. The resolved halo properties of the matched galaxies point in the same direction: although the velocity dispersions and dark matter densities are statistically consistent across all models (for the sample of matched galaxies), the SIDM halos exhibit significantly less radially biased velocity distributions (i.e., lower velocity anisotropy). Velocity anisotropy modifies the bar-halo resonances \citep{Gherghinescu2026} and also affects bar formation and the final bar strength \citep{sellwood2016}. Both effects favor more efficient angular momentum exchange, which makes the velocity anisotropy a potential explanation for the enhanced dynamical friction in SIDM. This interpretation agrees with \cite{fischer2024}, who reported enhanced dynamical friction for perturbers moving through a SIDM background with a constant cross-section, albeit in a different astrophysical context, and with \cite{bosch2026}, who showed that self-interactions broaden the bar-halo resonances and thereby enhance the transfer of angular momentum from the bar to the halo, in line with the stronger bars found in the self-interacting models. Nevertheless, dynamical friction alone cannot account for our results. Although the fractional slow-down is somewhat larger in SIDM1, this difference is modest compared with the systematic offset in pattern speed already present at the earliest redshifts probed. 

The lower pattern speeds in the SIDM models are therefore already established at the earliest epochs at which bars are identified in our sample, suggesting that the origin of the difference is closely connected to bar formation. Our analysis at $z\leq2$ indicates that the modest enhancement of dynamical friction in SIDM1 plays only a secondary role in producing slow bars at low redshift. Our conclusions are based on a limited number of matched galaxies, ranging from 5 at $z=2$ to 20 at $z=0.2$, drawn from a single cosmological volume and a single galaxy formation model. Notably, even when a larger sample of $N_{\rm rel}$ galaxies is considered, bars remain significantly slower in the SIDM models (Figures \ref{fig5p} and \ref{fig5}).

Future studies extending the analysis to higher redshifts and directly identifying the epoch of bar formation for a statistical sample of galaxies will be essential for determining why bars acquire lower pattern speeds in SIDM models. Our results indicate that, for the self-interaction models considered here, SIDM does not alleviate the fast-bar tension. Instead, it further sharpens the discrepancy and reinforces the potential of galactic bar dynamics as a sensitive probe of the microscopic nature of dark matter. Nevertheless, care is needed when interpreting the slowdown of bars over time. Because the galaxy sample changes between redshifts, a lower average pattern speed does not necessarily imply that the same bars have slowed down. A complementary study could instead follow a small sample of galaxies that remain barred at all redshifts and compare them on a case-by-case basis among the different dark matter models. As a final remark, higher-resolution simulations are necessary to draw more reliable conclusions about bar dynamics. As shown by the comparison between TNG50-1 and TNG50-2 simulations, numerical resolution can substantially affect the overall behaviour and properties of stellar bars \citep{2022ApJ...940...61F}.

\begin{acknowledgements}
We thank Dylan Nelson and Annalisa Pillepich for reading the manuscript and for their helpful comments. The work of MR and AH is supported by the Ferdowsi University of Mashhad. This work is based upon research funded by Iran National Science Foundation (INSF) under project No. 4046413. GD acknowledges the EuroHPC Joint Undertaking, ISCRA and ICSC for awarding the AIDA-TNG project access to the EuroHPC supercomputers LUMI and LEONARDO. 
\end{acknowledgements}

\bibliographystyle{aa}
\bibliography{DF_SIDM}

@ARTICLE{erwin2018,
       author = {{Erwin}, Peter},
        title = "{The dependence of bar frequency on galaxy mass, colour, and gas content - and angular resolution - in the local universe}",
      journal = {\mnras},
         year = 2018,
        month = mar,
       volume = {474},
       number = {4},
        pages = {5372-5392},
          doi = {10.1093/mnras/stx3117},
archivePrefix = {arXiv},
       eprint = {1711.04867},
 primaryClass = {astro-ph.GA},
       adsurl = {https://ui.adsabs.harvard.edu/abs/2018MNRAS.474.5372E}}

@INPROCEEDINGS{2009ASSP....8..159G,
       author = {{Gadotti}, Dimitri A.},
        title = "{Barred Galaxies: an Observer's Perspective}",
    booktitle = {Chaos in Astronomy},
         year = 2009,
       editor = {{Contopoulos}, G. and {Patsis}, P.~A.},
       series = {Astrophysics and Space Science Proceedings},
       volume = {8},
        month = jan,
        pages = {159},
          doi = {10.1007/978-3-540-75826-6_15},
archivePrefix = {arXiv},
       eprint = {0802.0495},
 primaryClass = {astro-ph},
       adsurl = {https://ui.adsabs.harvard.edu/abs/2009ASSP....8..159G}
}

@ARTICLE{math2025,
       author = {{Frosst}, Matthew and {Obreschkow}, Danail and {Ludlow}, Aaron and {Bottrell}, Connor and {Genel}, Shy},
        title = "{The complex relationship between black hole feedback, star formation, and stellar bars in TNG50}",
      journal = {\mnras},
        year = 2025,
        volume = {537},
        number = {4},
        pages = {3543--3552},
        month = feb,
          doi = {10.1093/mnras/staf255},
       adsurl = {https://ui.adsabs.harvard.edu/abs/2025MNRAS.tmp..256F}
}

@ARTICLE{at1992,
       author = {{Athanassoula}, E.},
        title = "{Morphology of bar orbits.}",
      journal = {\mnras},
         year = 1992,
        month = nov,
       volume = {259},
        pages = {328-344},
          doi = {10.1093/mnras/259.2.328},
       adsurl = {https://ui.adsabs.harvard.edu/abs/1992MNRAS.259..328A}
}

@ARTICLE{Athanassoula2013,
       author = {{Athanassoula}, E. and {Machado}, Rubens E.~G. and {Rodionov}, S.~A.},
        title = "{Bar formation and evolution in disc galaxies with gas and a triaxial halo: morphology, bar strength and halo properties}",
      journal = {\mnras},
         year = 2013,
        month = mar,
       volume = {429},
       number = {3},
        pages = {1949-1969},
          doi = {10.1093/mnras/sts452},
archivePrefix = {arXiv},
       eprint = {1211.6754},
 primaryClass = {astro-ph.CO},
       adsurl = {https://ui.adsabs.harvard.edu/abs/2013MNRAS.429.1949A}
}

@ARTICLE{2014RvMP...86....1S,
       author = {{Sellwood}, J.~A.},
        title = "{Secular evolution in disk galaxies}",
      journal = {Reviews of Modern Physics},
         year = 2014,
        month = jan,
       volume = {86},
       number = {1},
        pages = {1-46},
          doi = {10.1103/RevModPhys.86.1},
archivePrefix = {arXiv},
       eprint = {1310.0403},
 primaryClass = {astro-ph.GA},
       adsurl = {https://ui.adsabs.harvard.edu/abs/2014RvMP...86....1S}
}

@ARTICLE{eagle,
       author = {{Schaye}, Joop and {Crain}, Robert A. and {Bower}, Richard G. and {Furlong}, Michelle and {Schaller}, Matthieu and {Theuns}, Tom and {Dalla Vecchia}, Claudio and {Frenk}, Carlos S. and {McCarthy}, I.~G. and {Helly}, John C. and {Jenkins}, Adrian and {Rosas-Guevara}, Y.~M. and {White}, Simon D.~M. and {Baes}, Maarten and {Booth}, C.~M. and {Camps}, Peter and {Navarro}, Julio F. and {Qu}, Yan and {Rahmati}, Alireza and {Sawala}, Till and {Thomas}, Peter A. and {Trayford}, James},
        title = "{The EAGLE project: simulating the evolution and assembly of galaxies and their environments}",
      journal = {\mnras},
         year = 2015,
        month = jan,
       volume = {446},
       number = {1},
        pages = {521-554},
          doi = {10.1093/mnras/stu2058},
archivePrefix = {arXiv},
       eprint = {1407.7040},
 primaryClass = {astro-ph.GA},
       adsurl = {https://ui.adsabs.harvard.edu/abs/2015MNRAS.446..521S}
}

@ARTICLE{tng1,
       author = {{Pillepich}, Annalisa and {Nelson}, Dylan and {Hernquist}, Lars and {Springel}, Volker and {Pakmor}, R{\"u}diger and {Torrey}, Paul and {Weinberger}, Rainer and {Genel}, Shy and {Naiman}, Jill P. and {Marinacci}, Federico and {Vogelsberger}, Mark},
        title = "{First results from the IllustrisTNG simulations: the stellar mass content of groups and clusters of galaxies}",
      journal = {\mnras},
         year = 2018,
        month = mar,
       volume = {475},
       number = {1},
        pages = {648-675},
          doi = {10.1093/mnras/stx3112},
archivePrefix = {arXiv},
       eprint = {1707.03406},
 primaryClass = {astro-ph.GA},
       adsurl = {https://ui.adsabs.harvard.edu/abs/2018MNRAS.475..648P}
}

@ARTICLE{tng2,
       author = {{Springel}, Volker and {Pakmor}, R{\"u}diger and {Pillepich}, Annalisa and {Weinberger}, Rainer and {Nelson}, Dylan and {Hernquist}, Lars and {Vogelsberger}, Mark and {Genel}, Shy and {Torrey}, Paul and {Marinacci}, Federico and {Naiman}, Jill},
        title = "{First results from the IllustrisTNG simulations: matter and galaxy clustering}",
      journal = {\mnras},
         year = 2018,
        month = mar,
       volume = {475},
       number = {1},
        pages = {676-698},
          doi = {10.1093/mnras/stx3304},
archivePrefix = {arXiv},
       eprint = {1707.03397},
 primaryClass = {astro-ph.GA},
       adsurl = {https://ui.adsabs.harvard.edu/abs/2018MNRAS.475..676S}
}

@ARTICLE{tng3,
       author = {{Nelson}, Dylan and {Pillepich}, Annalisa and {Springel}, Volker and {Weinberger}, Rainer and {Hernquist}, Lars and {Pakmor}, R{\"u}diger and {Genel}, Shy and {Torrey}, Paul and {Vogelsberger}, Mark and {Kauffmann}, Guinevere and {Marinacci}, Federico and {Naiman}, Jill},
        title = "{First results from the IllustrisTNG simulations: the galaxy colour bimodality}",
      journal = {\mnras},
         year = 2018,
        month = mar,
       volume = {475},
       number = {1},
        pages = {624-647},
          doi = {10.1093/mnras/stx3040},
archivePrefix = {arXiv},
       eprint = {1707.03395},
 primaryClass = {astro-ph.GA},
       adsurl = {https://ui.adsabs.harvard.edu/abs/2018MNRAS.475..624N}
}

@ARTICLE{tng4,
       author = {{Naiman}, Jill P. and {Pillepich}, Annalisa and {Springel}, Volker and {Ramirez-Ruiz}, Enrico and {Torrey}, Paul and {Vogelsberger}, Mark and {Pakmor}, R{\"u}diger and {Nelson}, Dylan and {Marinacci}, Federico and {Hernquist}, Lars and {Weinberger}, Rainer and {Genel}, Shy},
        title = "{First results from the IllustrisTNG simulations: a tale of two elements - chemical evolution of magnesium and europium}",
      journal = {\mnras},
         year = 2018,
        month = jun,
       volume = {477},
       number = {1},
        pages = {1206-1224},
          doi = {10.1093/mnras/sty618},
archivePrefix = {arXiv},
       eprint = {1707.03401},
 primaryClass = {astro-ph.GA},
       adsurl = {https://ui.adsabs.harvard.edu/abs/2018MNRAS.477.1206N}
}

@ARTICLE{tng5,
       author = {{Marinacci}, Federico and {Vogelsberger}, Mark and {Pakmor}, R{\"u}diger and {Torrey}, Paul and {Springel}, Volker and {Hernquist}, Lars and {Nelson}, Dylan and {Weinberger}, Rainer and {Pillepich}, Annalisa and {Naiman}, Jill and {Genel}, Shy},
        title = "{First results from the IllustrisTNG simulations: radio haloes and magnetic fields}",
      journal = {\mnras},
         year = 2018,
        month = nov,
       volume = {480},
       number = {4},
        pages = {5113-5139},
          doi = {10.1093/mnras/sty2206},
archivePrefix = {arXiv},
       eprint = {1707.03396},
 primaryClass = {astro-ph.CO},
       adsurl = {https://ui.adsabs.harvard.edu/abs/2018MNRAS.480.5113M}
}

@ARTICLE{TNG502,
       author = {{Nelson}, Dylan and {Pillepich}, Annalisa and {Springel}, Volker and {Pakmor}, R{\"u}diger and {Weinberger}, Rainer and {Genel}, Shy and {Torrey}, Paul and {Vogelsberger}, Mark and {Marinacci}, Federico and {Hernquist}, Lars},
        title = "{First results from the TNG50 simulation: galactic outflows driven by supernovae and black hole feedback}",
      journal = {\mnras},
         year = 2019,
        month = dec,
       volume = {490},
       number = {3},
        pages = {3234-3261},
          doi = {10.1093/mnras/stz2306},
archivePrefix = {arXiv},
       eprint = {1902.05554},
 primaryClass = {astro-ph.GA},
       adsurl = {https://ui.adsabs.harvard.edu/abs/2019MNRAS.490.3234N}
}

@ARTICLE{TNG501,
       author = {{Nelson}, Dylan and {Springel}, Volker and {Pillepich}, Annalisa and {Rodriguez-Gomez}, Vicente and {Torrey}, Paul and {Genel}, Shy and {Vogelsberger}, Mark and {Pakmor}, Ruediger and {Marinacci}, Federico and {Weinberger}, Rainer and {Kelley}, Luke and {Lovell}, Mark and {Diemer}, Benedikt and {Hernquist}, Lars},
        title = "{The IllustrisTNG simulations: public data release}",
      journal = {Computational Astrophysics and Cosmology},
         year = 2019,
        month = may,
       volume = {6},
       number = {1},
          eid = {2},
        pages = {2},
          doi = {10.1186/s40668-019-0028-x},
archivePrefix = {arXiv},
       eprint = {1812.05609},
 primaryClass = {astro-ph.GA},
       adsurl = {https://ui.adsabs.harvard.edu/abs/2019ComAC...6....2N}
}

@ARTICLE{TNG503,
    author = {{Pillepich}, Annalisa and {Nelson}, Dylan and {Springel}, Volker and {Pakmor}, R{\"u}diger and {Torrey}, Paul and {Weinberger}, Rainer and {Vogelsberger}, Mark and {Marinacci}, Federico and {Genel}, Shy and {van der Wel}, Arjen and {Hernquist}, Lars},
    title = "{First results from the TNG50 simulation: the evolution of stellar and gaseous discs across cosmic time}",
    journal = {\mnras},
    year = 2019,
    month = dec,
    volume = {490},
    number = {3},
    pages = {3196-3233},
    doi = {10.1093/mnras/stz2338},
    archivePrefix = {arXiv},
    eprint = {1902.05553},
    primaryClass = {astro-ph.GA},
    adsurl = {https://ui.adsabs.harvard.edu/abs/2019MNRAS.490.3196P}}

@ARTICLE{Aguerri2015,
       author = {{Aguerri}, J.~A.~L. and {M{\'e}ndez-Abreu}, J. and {Falc{\'o}n-Barroso}, J. and {Amorin}, A. and {Barrera-Ballesteros}, J. and {Cid Fernandes}, R. and {Garc{\'\i}a-Benito}, R. and {Garc{\'\i}a-Lorenzo}, B. and {Gonz{\'a}lez Delgado}, R.~M. and {Husemann}, B. and {Kalinova}, V. and {Lyubenova}, M. and {Marino}, R.~A. and {M{\'a}rquez}, I. and {Mast}, D. and {P{\'e}rez}, E. and {S{\'a}nchez}, S.~F. and {van de Ven}, G. and {Walcher}, C.~J. and {Backsmann}, N. and {Cortijo-Ferrero}, C. and {Bland-Hawthorn}, J. and {del Olmo}, A. and {Iglesias-P{\'a}ramo}, J. and {P{\'e}rez}, I. and {S{\'a}nchez-Bl{\'a}zquez}, P. and {Wisotzki}, L. and {Ziegler}, B.},
        title = "{Bar pattern speeds in CALIFA galaxies. I. Fast bars across the Hubble sequence}",
      journal = {\aap},
         year = 2015,
        month = apr,
       volume = {576},
          eid = {A102},
        pages = {A102},
          doi = {10.1051/0004-6361/201423383},
archivePrefix = {arXiv},
       eprint = {1501.05498},
 primaryClass = {astro-ph.GA},
       adsurl = {https://ui.adsabs.harvard.edu/abs/2015A&A...576A.102A}
}

@ARTICLE{algorry2017,
       author = {{Algorry}, David G. and {Navarro}, Julio F. and {Abadi}, Mario G. and {Sales}, Laura V. and {Bower}, Richard G. and {Crain}, Robert A. and {Dalla Vecchia}, Claudio and {Frenk}, Carlos S. and {Schaller}, Matthieu and {Schaye}, Joop and {Theuns}, Tom},
        title = "{Barred galaxies in the EAGLE cosmological hydrodynamical simulation}",
      journal = {\mnras},
         year = 2017,
        month = jul,
       volume = {469},
       number = {1},
        pages = {1054-1064},
          doi = {10.1093/mnras/stx1008},
archivePrefix = {arXiv},
       eprint = {1609.05909},
 primaryClass = {astro-ph.GA},
       adsurl = {https://ui.adsabs.harvard.edu/abs/2017MNRAS.469.1054A}
}

@ARTICLE{Cuomo2020,
       author = {{Cuomo}, V. and {Aguerri}, J.~A.~L. and {Corsini}, E.~M. and {Debattista}, V.~P.},
        title = "{Relations among structural parameters in barred galaxies with a direct measurement of bar pattern speed}",
      journal = {\aap},
         year = 2020,
        month = sep,
       volume = {641},
          eid = {A111},
        pages = {A111},
          doi = {10.1051/0004-6361/202037945},
archivePrefix = {arXiv},
       eprint = {2003.07455},
 primaryClass = {astro-ph.GA},
       adsurl = {https://ui.adsabs.harvard.edu/abs/2020A&A...641A.111C}
}

@ARTICLE{2021MNRAS.508..926R,
       author = {{Roshan}, Mahmood and {Ghafourian}, Neda and {Kashfi}, Tahere and {Banik}, Indranil and {Haslbauer}, Moritz and {Cuomo}, Virginia and {Famaey}, Benoit and {Kroupa}, Pavel},
        title = "{Fast galaxy bars continue to challenge standard cosmology}",
      journal = {\mnras},
         year = 2021,
        month = nov,
       volume = {508},
       number = {1},
        pages = {926-939},
          doi = {10.1093/mnras/stab2553},
archivePrefix = {arXiv},
       eprint = {2106.10304},
 primaryClass = {astro-ph.GA},
       adsurl = {https://ui.adsabs.harvard.edu/abs/2021MNRAS.508..926R}
}

@ARTICLE{Geron2023,
       author = {{G{\'e}ron}, Tobias and {Smethurst}, Rebecca J. and {Lintott}, Chris and {Kruk}, Sandor and {Masters}, Karen L. and {Simmons}, Brooke and {Mantha}, Kameswara Bharadwaj and {Walmsley}, Mike and {Garma-Oehmichen}, L. and {Drory}, Niv and {Lane}, Richard R.},
        title = "{Galaxy Zoo: kinematics of strongly and weakly barred galaxies}",
      journal = {\mnras},
         year = 2023,
        month = may,
       volume = {521},
       number = {2},
        pages = {1775-1793},
          doi = {10.1093/mnras/stad501},
archivePrefix = {arXiv},
       eprint = {2302.05464},
 primaryClass = {astro-ph.GA},
       adsurl = {https://ui.adsabs.harvard.edu/abs/2023MNRAS.521.1775G}
}

@ARTICLE{tw1984,
   author = {{Tremaine}, S. and {Weinberg}, M.~D.},
    title = {Dynamical friction in spherical systems},
  journal = {\mnras},
     year = 1984,
   volume = {209},
    pages = {729},
      doi = {10.1093/mnras/209.4.729}
}

@ARTICLE{weinberg1985,
   author = {{Weinberg}, M.~D.},
    title = {Evolution of barred galaxies by dynamical friction},
  journal = {\mnras},
     year = 1985,
   volume = {213},
    pages = {451},
      doi = {10.1093/mnras/213.3.451}
}

@ARTICLE{debattista2000,
       author = {{Debattista}, Victor P. and {Sellwood}, J.~A.},
        title = "{Constraints from Dynamical Friction on the Dark Matter Content of Barred Galaxies}",
      journal = {\apj},
         year = 2000,
        month = nov,
       volume = {543},
       number = {2},
        pages = {704-721},
          doi = {10.1086/317148},
       adsurl = {https://ui.adsabs.harvard.edu/abs/2000ApJ...543..704D}}

@ARTICLE{athanassoula2003,
   author = {{Athanassoula}, E.},
    title = {What determines the strength and the slowdown rate of bars?},
  journal = {\mnras},
     year = 2003,
   volume = {341},
    pages = {1179},
      doi = {10.1046/j.1365-8711.2003.06473.x}
}

@ARTICLE{Fragkoudi2021,
       author = {{Fragkoudi}, F. and {Grand}, R.~J.~J. and {Pakmor}, R. and {Springel}, V. and {White}, S.~D.~M. and {Marinacci}, F. and {Gomez}, F.~A. and {Navarro}, J.~F.},
        title = "{Revisiting the tension between fast bars and the {\ensuremath{\Lambda}}CDM paradigm}",
      journal = {\aap},
         year = 2021,
        month = jun,
       volume = {650},
          eid = {L16},
        pages = {L16},
          doi = {10.1051/0004-6361/202140320},
archivePrefix = {arXiv},
       eprint = {2011.13942},
 primaryClass = {astro-ph.GA},
       adsurl = {https://ui.adsabs.harvard.edu/abs/2021A&A...650L..16F}
}

@ARTICLE{hui2017,
   author = {{Hui}, Lam and {Ostriker}, Jeremiah P. and {Tremaine}, Scott and {Witten}, Edward},
    title = {Ultralight scalars as cosmological dark matter},
  journal = {\prd},
     year = 2017,
   volume = {95},
    pages = {043541},
      doi = {10.1103/PhysRevD.95.043541}
}

@ARTICLE{lancaster2020,
   author = {{Lancaster}, Lachlan and {Giovanetti}, Cara and {Mocz}, Philip and {Kahn}, Yonatan and {Lisanti}, Mariangela and {Spergel}, David N.},
    title = {Dynamical friction in a Fuzzy Dark Matter universe},
  journal = {\jcap},
     year = 2020,
   volume = {2020},
   number = {1},
    pages = {001},
      doi = {10.1088/1475-7516/2020/01/001}
}

@ARTICLE{spergel2000,
   author = {{Spergel}, David N. and {Steinhardt}, Paul J.},
    title = {Observational Evidence for Self-Interacting Cold Dark Matter},
  journal = {\prl},
     year = 2000,
   volume = {84},
    pages = {3760},
      doi = {10.1103/PhysRevLett.84.3760}
}

@ARTICLE{tulin2018,
   author = {{Tulin}, Sean and {Yu}, Hai-Bo},
    title = {Dark matter self-interactions and small scale structure},
  journal = {\physrep},
     year = 2018,
   volume = {730},
    pages = {1},
      doi = {10.1016/j.physrep.2017.11.004}
}

@ARTICLE{adhikari2022,
       author = {{Adhikari}, Susmita and {Banerjee}, Arka and {Boddy}, Kimberly K. and {Cyr-Racine}, Francis-Yan and {Desmond}, Harry and {Dvorkin}, Cora and {Jain}, Bhuvnesh and {Kahlhoefer}, Felix and {Kaplinghat}, Manoj and {Nierenberg}, Anna and {Peter}, Annika H.~G. and {Robertson}, Andrew and {Sakstein}, Jeremy and {Zavala}, Jes{\'u}s},
        title = "{Astrophysical tests of dark matter self-interactions}",
      journal = {Reviews of Modern Physics},
         year = 2025,
        month = oct,
       volume = {97},
       number = {4},
          eid = {045004},
        pages = {045004},
          doi = {10.1103/m2vm-59y3},
archivePrefix = {arXiv},
       eprint = {2207.10638},
 primaryClass = {astro-ph.CO},
       adsurl = {https://ui.adsabs.harvard.edu/abs/2025RvMP...97d5004A}
}

@ARTICLE{deblok2010,
   author = {{de Blok}, W.~J.~G.},
    title = {The Core-Cusp Problem},
  journal = {Advances in Astronomy},
     year = 2010,
   volume = {2010},
    pages = {789293},
      doi = {10.1155/2010/789293}
}

@ARTICLE{despali2026,
   author = {{Despali}, Giulia and {Giocoli}, Carlo and {Moscardini}, Lauro and {Pillepich}, Annalisa and {Vogelsberger}, Mark and {Meneghetti}, Massimo},
    title = {The AIDA-TNG project: Dark matter profiles and concentrations in alternative dark matter models},
  journal = {\aap},
     year = 2026,
   volume = {708},
    pages = {A47},
      doi = {10.1051/0004-6361/202558686}
}

@ARTICLE{fischer2024,
       author = {{Fischer}, Moritz S. and {Sagunski}, Laura},
        title = "{Dynamical friction from self-interacting dark matter}",
      journal = {\aap},
         year = 2024,
        month = oct,
       volume = {690},
          eid = {A299},
        pages = {A299},
          doi = {10.1051/0004-6361/202451304},
archivePrefix = {arXiv},
       eprint = {2405.19392},
 primaryClass = {astro-ph.CO},
       adsurl = {https://ui.adsabs.harvard.edu/abs/2024A&A...690A.299F}
}

@ARTICLE{bosch2026,
       author = {{Dattathri}, Shashank and {van den Bosch}, Frank C. and {Zhang}, HanYuan and {Weinberg}, Martin D. and {Vasiliev}, Eugene and {Natarajan}, Priyamvada and {Belokurov}, Vasily},
        title = "{Self-interacting dark matter promotes bar formation in disk galaxies}",
      journal = {\mnras},
         year = 2026,
         volume = {552},
        number = {4},
        pages = {stag1847},
        month = sep,
          doi = {10.1093/mnras/stag1847},
       adsurl = {https://ui.adsabs.harvard.edu/abs/2026MNRAS.tmp.1706D}
}

@ARTICLE{guo2023,
   author = {{Guo}, Yuchen and others},
    title = {First Look at z > 1 Bars in the Rest-frame Near-infrared with JWST Early CEERS Imaging},
  journal = {\apjl},
     year = 2023,
   volume = {945},
    pages = {L10},
      doi = {10.3847/2041-8213/acacfb},
}

@ARTICLE{leconte2024,
    author = {{Le Conte}, Zoe A. and others},
    title = {A JWST investigation into the bar fraction at redshifts 1 < z < 3},
  journal = {\mnras},
     year = 2024,
   volume = {530},
    pages = {1984},
      doi = {10.1093/mnras/stae921},

}

@ARTICLE{despali2025,
       author = {{Despali}, Giulia and {Moscardini}, Lauro and {Nelson}, Dylan and {Pillepich}, Annalisa and {Springel}, Volker and {Vogelsberger}, Mark},
        title = "{Introducing the AIDA-TNG project: Galaxy formation in alternative dark matter models}",
      journal = {\aap},
         year = 2025,
        month = may,
       volume = {697},
          eid = {A213},
        pages = {A213},
          doi = {10.1051/0004-6361/202553836},
archivePrefix = {arXiv},
       eprint = {2501.12439},
 primaryClass = {astro-ph.CO},
       adsurl = {https://ui.adsabs.harvard.edu/abs/2025A&A...697A.213D}
}

@ARTICLE{sales2010,
       author = {{Sales}, Laura V. and {Navarro}, Julio F. and {Schaye}, Joop and {Dalla Vecchia}, Claudio and {Springel}, Volker and {Booth}, C.~M.},
        title = "{Feedback and the structure of simulated galaxies at redshift z= 2}",
      journal = {\mnras},
         year = 2010,
        month = dec,
       volume = {409},
       number = {4},
        pages = {1541-1556},
          doi = {10.1111/j.1365-2966.2010.17391.x},
archivePrefix = {arXiv},
       eprint = {1004.5386},
 primaryClass = {astro-ph.CO},
       adsurl = {https://ui.adsabs.harvard.edu/abs/2010MNRAS.409.1541S}
}

@ARTICLE{genel2015,
       author = {{Genel}, Shy and {Fall}, S. Michael and {Hernquist}, Lars and {Vogelsberger}, Mark and {Snyder}, Gregory F. and {Rodriguez-Gomez}, Vicente and {Sijacki}, Debora and {Springel}, Volker},
        title = "{Galactic Angular Momentum in the Illustris Simulation: Feedback and the Hubble Sequence}",
      journal = {\apjl},
         year = 2015,
        month = may,
       volume = {804},
       number = {2},
          eid = {L40},
        pages = {L40},
          doi = {10.1088/2041-8205/804/2/L40},
archivePrefix = {arXiv},
       eprint = {1503.01117},
 primaryClass = {astro-ph.GA},
       adsurl = {https://ui.adsabs.harvard.edu/abs/2015ApJ...804L..40G}
}

@ARTICLE{ohta,
       author = {{Ohta}, Kouji and {Hamabe}, Masaru and {Wakamatsu}, Ken-Ichi},
        title = "{Surface Photometry of Barred Spiral Galaxies}",
      journal = {\apj},
         year = 1990,
        month = jul,
       volume = {357},
        pages = {71},
          doi = {10.1086/168892},
       adsurl = {https://ui.adsabs.harvard.edu/abs/1990ApJ...357...71O}
}

@ARTICLE{Aguerri_2000,
       author = {{Aguerri}, J.~A.~L. and {Mu{\~n}oz-Tu{\~n}{\'o}n}, C. and {Varela}, A.~M. and {Prieto}, M.},
        title = "{Characterizing bar structures: application to NGC 1300, NGC 7479 and NGC 7723}",
      journal = {A\&A},
         year = 2000,
        month = sep,
       volume = {361},
        pages = {841-849},
       adsurl = {https://ui.adsabs.harvard.edu/abs/2000A\&A...361..841A}}

@ARTICLE{Laurikainen2008,
       author = {{Das}, M. and {Laurikainen}, E. and {Salo}, H. and {Buta}, R.},
        title = "{Variation of bar strength with central velocity dispersion in spiral galaxies}",
      journal = {\apss},
         year = 2008,
        month = oct,
       volume = {317},
       number = {3-4},
        pages = {163-168},
          doi = {10.1007/s10509-008-9873-9},
archivePrefix = {arXiv},
       eprint = {0807.2699},
 primaryClass = {astro-ph},
       adsurl = {https://ui.adsabs.harvard.edu/abs/2008Ap&SS.317..163D}
}

@ARTICLE{roshan2025,
       author = {{Roshan}, Mahmood and {Habibi}, Asiyeh and {Aguerri}, J. Alfonso L. and {Cuomo}, Virginia and {Bottrell}, Connor and {Costantin}, Luca and {Corsini}, Enrico Maria and {Kim}, Taehyun and {Lee}, Yun Hee and {Mendez-Abreu}, Jairo and {Frosst}, Matthew and {de Lorenzo-C{\'a}ceres}, Adriana and {Morelli}, Lorenzo and {Pizzella}, Alessandro},
        title = "{The Tremaine-Weinberg method at high redshifts}",
      journal = {\aap},
         year = 2025,
        month = sep,
       volume = {701},
          eid = {A160},
        pages = {A160},
          doi = {10.1051/0004-6361/202555547},
archivePrefix = {arXiv},
       eprint = {2507.18292},
 primaryClass = {astro-ph.GA},
       adsurl = {https://ui.adsabs.harvard.edu/abs/2025A&A...701A.160R}
}

@ARTICLE{dehnen2024,
       author = {{Semczuk}, Marcin and {Dehnen}, Walter and {Sch{\"o}nrich}, Ralph and {Athanassoula}, E.},
        title = "{Pattern speed evolution of barred galaxies in TNG50}",
      journal = {\aap},
         year = 2024,
        month = dec,
       volume = {692},
          eid = {A159},
        pages = {A159},
          doi = {10.1051/0004-6361/202451521},
archivePrefix = {arXiv},
       eprint = {2407.11154},
 primaryClass = {astro-ph.GA},
       adsurl = {https://ui.adsabs.harvard.edu/abs/2024A&A...692A.159S}
}

@ARTICLE{zhang2024,
       author = {{Zhang}, Hanyuan and {Belokurov}, Vasily and {Evans}, N. Wyn and {Kane}, Sarah G. and {Sanders}, Jason L.},
        title = "{Kinematics and dynamics of the Galactic bar revealed by Gaia long-period variables}",
      journal = {\mnras},
         year = 2024,
        month = sep,
       volume = {533},
       number = {3},
        pages = {3395-3414},
          doi = {10.1093/mnras/stae2023},
archivePrefix = {arXiv},
       eprint = {2406.06678},
 primaryClass = {astro-ph.GA},
       adsurl = {https://ui.adsabs.harvard.edu/abs/2024MNRAS.533.3395Z}
}

@ARTICLE{Arranz2024,
       author = {{Jim{\'e}nez-Arranz}, {\'O}. and {Chemin}, L. and {Romero-G{\'o}mez}, M. and {Luri}, X. and {Adamczyk}, P. and {Castro-Ginard}, A. and {Roca-F{\`a}brega}, S. and {McMillan}, P.~J. and {Cioni}, M. -R.~L.},
        title = "{The bar pattern speed of the Large Magellanic Cloud}",
      journal = {\aap},
         year = 2024,
        month = mar,
       volume = {683},
          eid = {A102},
        pages = {A102},
          doi = {10.1051/0004-6361/202347266},
archivePrefix = {arXiv},
       eprint = {2312.11192},
 primaryClass = {astro-ph.GA},
       adsurl = {https://ui.adsabs.harvard.edu/abs/2024A&A...683A.102J}
}

@ARTICLE{Rosas2022,
       author = {{Rosas-Guevara}, Yetli and {Bonoli}, Silvia and {Dotti}, Massimo and {Izquierdo-Villalba}, David and {Lupi}, Alessandro and {Zana}, Tommaso and {Bonetti}, Matteo and {Nelson}, Dylan and {Springel}, Volker and {Hernquist}, Lars and {Vogelsberger}, Mark},
        title = "{The evolution of the barred galaxy population in the TNG50 simulation}",
      journal = {\mnras},
         year = 2022,
        month = jun,
       volume = {512},
       number = {4},
        pages = {5339-5357},
          doi = {10.1093/mnras/stac816},
archivePrefix = {arXiv},
       eprint = {2110.04537},
 primaryClass = {astro-ph.GA},
       adsurl = {https://ui.adsabs.harvard.edu/abs/2022MNRAS.512.5339R}
}

@ARTICLE{AHabibi,
       author = {{Habibi}, Asiyeh and {Roshan}, Mahmood and {Hosseinirad}, Mohammad and {Khosroshahi}, Habib and {Aguerri}, J.~A.~L. and {Cuomo}, Virginia and {Abbassi}, Shahram},
        title = "{The redshift evolution of galactic bar pattern speed in TNG50}",
      journal = {\aap},
         year = 2024,
        month = nov,
       volume = {691},
          eid = {A122},
        pages = {A122},
          doi = {10.1051/0004-6361/202451028},
archivePrefix = {arXiv},
       eprint = {2409.02456},
 primaryClass = {astro-ph.GA},
       adsurl = {https://ui.adsabs.harvard.edu/abs/2024A&A...691A.122H}
}

@ARTICLE{2016A&A...587A.160D,
       author = {{D{\'\i}az-Garc{\'\i}a}, S. and {Salo}, H. and {Laurikainen}, E. and {Herrera-Endoqui}, M.},
        title = "{Characterization of galactic bars from 3.6 {\ensuremath{\mu}}m S$^{4}$G imaging}",
      journal = {\aap},
         year = 2016,
        month = mar,
       volume = {587},
          eid = {A160},
        pages = {A160},
          doi = {10.1051/0004-6361/201526161},
archivePrefix = {arXiv},
       eprint = {1509.06743},
 primaryClass = {astro-ph.GA},
       adsurl = {https://ui.adsabs.harvard.edu/abs/2016A&A...587A.160D}
}

@ARTICLE{2022ApJ...940...61F,
       author = {{Frankel}, Neige and {Pillepich}, Annalisa and {Rix}, Hans-Walter and {Rodriguez-Gomez}, Vicente and {Sanders}, Jason and {Bovy}, Jo and {Kollmeier}, Juna and {Murray}, Norm and {Mackereth}, Ted},
        title = "{Simulated Bars May Be Shorter but Are Not Slower Than Those Observed: TNG50 versus MaNGA}",
      journal = {\apj},
         year = 2022,
        month = nov,
       volume = {940},
       number = {1},
          eid = {61},
        pages = {61},
          doi = {10.3847/1538-4357/ac9972},
archivePrefix = {arXiv},
       eprint = {2201.08406},
 primaryClass = {astro-ph.GA},
       adsurl = {https://ui.adsabs.harvard.edu/abs/2022ApJ...940...61F}
}

@ARTICLE{binney1977,
       author = {{Binney}, J.},
        title = "{Dynamical friction in aspherical clusters.}",
      journal = {\mnras},
         year = 1977,
        month = dec,
       volume = {181},
        pages = {735-746},
          doi = {10.1093/mnras/181.4.735},
       adsurl = {https://ui.adsabs.harvard.edu/abs/1977MNRAS.181..735B}
}

@ARTICLE{kroupa2004,
       author = {{Pe{\~n}arrubia}, Jorge and {Just}, Andreas and {Kroupa}, Pavel},
        title = "{Dynamical friction in flattened systems: a numerical test of Binney's approach}",
      journal = {\mnras},
         year = 2004,
        month = apr,
       volume = {349},
       number = {2},
        pages = {747-756},
          doi = {10.1111/j.1365-2966.2004.07549.x},
archivePrefix = {arXiv},
       eprint = {astro-ph/0401159},
 primaryClass = {astro-ph},
       adsurl = {https://ui.adsabs.harvard.edu/abs/2004MNRAS.349..747P}
}

@ARTICLE{sellwood2006,
       author = {{Sellwood}, J.~A.},
        title = "{Bar-Halo Friction in Galaxies. I. Scaling Laws}",
      journal = {\apj},
         year = 2006,
        month = feb,
       volume = {637},
       number = {2},
        pages = {567-581},
          doi = {10.1086/498418},
archivePrefix = {arXiv},
       eprint = {astro-ph/0407533},
 primaryClass = {astro-ph},
       adsurl = {https://ui.adsabs.harvard.edu/abs/2006ApJ...637..567S}
}

@ARTICLE{sellwood2016,
       author = {{Sellwood}, J.~A.},
        title = "{Bar Instability in Disk-Halo Systems}",
      journal = {\apj},
         year = 2016,
        month = mar,
       volume = {819},
       number = {2},
          eid = {92},
        pages = {92},
          doi = {10.3847/0004-637X/819/2/92},
archivePrefix = {arXiv},
       eprint = {1601.03406},
 primaryClass = {astro-ph.GA},
       adsurl = {https://ui.adsabs.harvard.edu/abs/2016ApJ...819...92S}
}

@ARTICLE{Gherghinescu2026,
       author = {{Gherghinescu},Paula and {Fragkoudi}, Francesca and {Deason}, Alis},
        title = "{Bar-halo interaction: the role of orbital anisotropy}",
      journal = {arXiv e-prints},
         year = 2026,
        month = may,
          eid = {arXiv:2605.26225},
        pages = {arXiv:2605.26225},
          doi = {10.48550/arXiv.2605.26225},
archivePrefix = {arXiv},
       eprint = {2605.26225},
 primaryClass = {astro-ph.GA},
       adsurl = {https://ui.adsabs.harvard.edu/abs/2026arXiv260526225G}
}

@ARTICLE{1972MNRAS.157....1L,
       author = {{Lynden-Bell}, D. and {Kalnajs}, A.~J.},
        title = "{On the generating mechanism of spiral structure}",
      journal = {\mnras},
         year = 1972,
        month = jan,
       volume = {157},
        pages = {1},
          doi = {10.1093/mnras/157.1.1},
       adsurl = {https://ui.adsabs.harvard.edu/abs/1972MNRAS.157....1L}
}

@INPROCEEDINGS{fire2,
       author = {{Ansar}, Sioree and {Sanderson}, Robyn and {Pearson}, Sarah},
        title = "{Bar Formation and Destruction in the FIRE2 Simulations}",
    booktitle = {AAS/Division of Dynamical Astronomy Meeting},
         year = 2022,
       series = {AAS/Division of Dynamical Astronomy Meeting},
       volume = {54},
        month = may,
          eid = {400.02},
        pages = {400.02},
       adsurl = {https://ui.adsabs.harvard.edu/abs/2022DDA....5340002A}
}

@ARTICLE{Zavala2013,
  author  = {{Zavala}, Jes{\'u}s and {Vogelsberger}, Mark and {Walker}, Matthew G.},
  title   = "{Constraining self-interacting dark matter with the Milky Way's dwarf spheroidals}",
  journal = {\mnras},
  year    = {2013},
  volume  = {431},
  number  = {1},
  pages   = {L20--L24},
  doi     = {10.1093/mnrasl/sls053}
}

@ARTICLE{Vogelsberger2012,
  author  = {{Vogelsberger}, Mark and {Zavala}, Jes{\'u}s and {Loeb}, Abraham},
  title   = "{Subhaloes in self-interacting galactic dark matter haloes}",
  journal = {\mnras},
  year    = {2012},
  volume  = {423},
  number  = {4},
  pages   = {3740--3752},
  doi     = {10.1111/j.1365-2966.2012.21182.x}
}

@ARTICLE{Creasey2017,
  author  = {{Creasey}, Peter and {Sameie}, Omid and {Sales}, Laura V. and {Yu}, Hai-Bo and {Vogelsberger}, Mark and {Zavala}, Jes{\'u}s},
  title   = "{Spreading out and staying sharp -- creating diverse rotation curves via baryonic and self-interaction effects}",
  journal = {\mnras},
  year    = {2017},
  volume  = {468},
  number  = {2},
  pages   = {2283--2295},
  doi     = {10.1093/mnras/stx522}
}

@ARTICLE{Vogelsberger2019,
  author  = {{Vogelsberger}, Mark and {Zavala}, Jes{\'u}s and {Schutz}, Katelin and {Slatyer}, Tracy R.},
  title   = "{Evaporating the Milky Way halo and its satellites with inelastic self-interacting dark matter}",
  journal = {\mnras},
  year    = {2019},
  volume  = {484},
  number  = {4},
  pages   = {5437--5452},
  doi     = {10.1093/mnras/stz340}
}

@ARTICLE{Zeng2022,
  author  = {{Zeng}, Zhichao Carton and {Peter}, Annika H.~G. and {Du}, Xiaolong and {Benson}, Andrew and {Kim}, Stacy Y. and {Jiang}, Fangzhou and {Cyr-Racine}, Francis-Yan and {Vogelsberger}, Mark},
  title   = "{Core-collapse, evaporation, and tidal effects: the life story of a self-interacting dark matter subhalo}",
  journal = {\mnras},
  year    = {2022},
  volume  = {513},
  number  = {4},
  pages   = {4845--4868},
  doi     = {10.1093/mnras/stac1094}
}

@ARTICLE{Sameie2018,
  author  = {{Sameie}, Omid and {Creasey}, Peter and {Yu}, Hai-Bo and {Sales}, Laura V. and {Vogelsberger}, Mark and {Zavala}, Jes{\'u}s},
  title   = "{The impact of baryonic discs on the shapes and profiles of self-interacting dark matter haloes}",
  journal = {\mnras},
  year    = {2018},
  volume  = {479},
  number  = {1},
  pages   = {359--367},
  doi     = {10.1093/mnras/sty1516}
}

@ARTICLE{CyrRacine2016,
  author  = {{Cyr-Racine}, Francis-Yan and {Sigurdson}, Kris and {Zavala}, Jes{\'u}s and {Bringmann}, Torsten and {Vogelsberger}, Mark and {Pfrommer}, Christoph},
  title   = "{{ETHOS}---an effective theory of structure formation: from dark particle physics to the matter distribution of the Universe}",
  journal = {\prd},
  year    = {2016},
  volume  = {93},
  number  = {12},
  pages   = {123527},
  doi     = {10.1103/PhysRevD.93.123527}
}

@ARTICLE{Vogelsberger2016,
  author  = {{Vogelsberger}, Mark and {Zavala}, Jes{\'u}s and {Cyr-Racine}, Francis-Yan and {Pfrommer}, Christoph and {Bringmann}, Torsten and {Sigurdson}, Kris},
  title   = "{{ETHOS}---an effective theory of structure formation: dark matter physics as a possible explanation of the small-scale {CDM} problems}",
  journal = {\mnras},
  year    = {2016},
  volume  = {460},
  number  = {2},
  pages   = {1399--1416},
  doi     = {10.1093/mnras/stw1076}
}

@ARTICLE{Rose2023,
       author = {{Rose}, Jonah C. and {Torrey}, Paul and {Vogelsberger}, Mark and {O'Neil}, Stephanie},
        title = "{Unravelling the interplay between SIDM and baryons in MW haloes: defining where baryons dictate heat transfer}",
      journal = {\mnras},
         year = 2023,
        month = mar,
       volume = {519},
       number = {4},
        pages = {5623-5636},
          doi = {10.1093/mnras/stac3634},
archivePrefix = {arXiv},
       eprint = {2206.14830},
 primaryClass = {astro-ph.GA},
       adsurl = {https://ui.adsabs.harvard.edu/abs/2023MNRAS.519.5623R}
}

@ARTICLE{2024A&A...684A.179R,
       author = {{Rosas-Guevara}, Yetli and {Bonoli}, Silvia and {Misa Moreira}, Carmen and {Izquierdo-Villalba}, David},
        title = "{The rise and fall of bars in disc galaxies from z = 1 to z = 0. The role of environment}",
      journal = {\aap},
         year = 2024,
        month = apr,
       volume = {684},
          eid = {A179},
        pages = {A179},
          doi = {10.1051/0004-6361/202349003},
archivePrefix = {arXiv},
       eprint = {2401.15215},
 primaryClass = {astro-ph.GA},
       adsurl = {https://ui.adsabs.harvard.edu/abs/2024A&A...684A.179R}
}

@ARTICLE{2026MNRAS.546ag263G,
       author = {{Gon{\c{c}}alves}, Gustavo F. and {Machado}, Rubens E.~G. and {Valen{\c{c}}a}, Raquel R. and {Athanassoula}, E. and {Men{\'e}ndez-Delmestre}, Kar{\'\i}n and {Bueno-Dalpiaz}, Thiago},
        title = "{The detectability of bars at high redshift: a case study using Euclid-like mock observations of TNG50 simulated galaxies}",
      journal = {\mnras},
         year = 2026,
        month = mar,
       volume = {546},
       number = {4},
          eid = {stag263},
        pages = {stag263},
          doi = {10.1093/mnras/stag263},
archivePrefix = {arXiv},
       eprint = {2601.16335},
 primaryClass = {astro-ph.GA},
       adsurl = {https://ui.adsabs.harvard.edu/abs/2026MNRAS.546ag263G}
}

@ARTICLE{2025A&A...698A..20R,
       author = {{Rosas-Guevara}, Yetli and {Bonoli}, Silvia and {Puchwein}, Ewald and {Dotti}, Massimo and {Contreras}, Sergio},
        title = "{Galaxy formation physics behind bar formation: A view from cosmological hydrodynamical simulations}",
      journal = {\aap},
         year = 2025,
        month = jun,
       volume = {698},
          eid = {A20},
        pages = {A20},
          doi = {10.1051/0004-6361/202453160},
archivePrefix = {arXiv},
       eprint = {2411.16876},
 primaryClass = {astro-ph.GA},
       adsurl = {https://ui.adsabs.harvard.edu/abs/2025A&A...698A..20R}
}

@ARTICLE{2025MNRAS.542.3154G,
       author = {{Gon{\c{c}}alves}, Gustavo F. and {Machado}, Rubens E.~G. and {Men{\'e}ndez-Delmestre}, Kar{\'\i}n and {Bueno-Dalpiaz}, Thiago},
        title = "{Bar properties as a function of wavelength in TNG50: analysis of mock images}",
      journal = {\mnras},
         year = 2025,
        month = aug,
       volume = {542},
       number = {4},
        pages = {3154-1380},
          doi = {10.1093/mnras/staf1384},
archivePrefix = {arXiv},
       eprint = {2506.19373},
 primaryClass = {astro-ph.GA},
       adsurl = {https://ui.adsabs.harvard.edu/abs/2025MNRAS.542.3154G}
}

@ARTICLE{2026arXiv260912339C,
       author = {{Chim-Ramirez}, Karol and {Cervantes-Sodi}, Bernardo and {Rosas-Guevara}, Yetli},
        title = "{Stellar bar occurrence in TNG50 and TNG100: disentangling environmental effects and assembly history at z = 0}",
      journal = {arXiv e-prints},
         year = 2026,
        month = sep,
          eid = {arXiv:2609.12339},
        pages = {arXiv:2609.12339},
          doi = {10.48550/arXiv.2609.12339},
archivePrefix = {arXiv},
       eprint = {2609.12339},
 primaryClass = {astro-ph.GA},
       adsurl = {https://ui.adsabs.harvard.edu/abs/2026arXiv260912339C}
}

@ARTICLE{2022MNRAS.514.1006I,
       author = {{Izquierdo-Villalba}, David and {Bonoli}, Silvia and {Rosas-Guevara}, Yetli and {Springel}, Volker and {White}, Simon D.~M. and {Zana}, Tommaso and {Dotti}, Massimo and {Spinoso}, Daniele and {Bonetti}, Matteo and {Lupi}, Alessandro},
        title = "{Disc instability and bar formation: view from the IllustrisTNG simulations}",
      journal = {\mnras},
         year = 2022,
        month = jul,
       volume = {514},
       number = {1},
        pages = {1006-1020},
          doi = {10.1093/mnras/stac1413},
archivePrefix = {arXiv},
       eprint = {2203.07734},
 primaryClass = {astro-ph.GA},
       adsurl = {https://ui.adsabs.harvard.edu/abs/2022MNRAS.514.1006I}
}

@ARTICLE{2020ApJ...904..170Z,
       author = {{Zhao}, Dongyao and {Du}, Min and {Ho}, Luis C. and {Debattista}, Victor P. and {Shi}, Jingjing},
        title = "{Barred Galaxies in the IllustrisTNG Simulation}",
      journal = {\apj},
         year = 2020,
        month = dec,
       volume = {904},
       number = {2},
          eid = {170},
        pages = {170},
          doi = {10.3847/1538-4357/abbe1b},
archivePrefix = {arXiv},
       eprint = {2009.06895},
 primaryClass = {astro-ph.GA},
       adsurl = {https://ui.adsabs.harvard.edu/abs/2020ApJ...904..170Z}
}

@ARTICLE{Torrey14,
   author = {{Torrey}, P. and {Vogelsberger}, M. and {Genel}, S. and {Sijacki}, D. and 
	{Springel}, V. and {Hernquist}, L.},
    title = "{A model for cosmological simulations of galaxy formation physics: multi-epoch validation}",
  journal = {\mnras},
archivePrefix = "arXiv",
   eprint = {1305.4931},
     year = 2014,
    month = mar,
   volume = 438,
    pages = {1985-2004},
      doi = {10.1093/mnras/stt2295},
   adsurl = {http://adsabs.harvard.edu/abs/2014MNRAS.438.1985T}
}

@ARTICLE{weinberger17,
   author = {{Weinberger}, R. and {Springel}, V. and {Hernquist}, L. and 
	{Pillepich}, A. and {Marinacci}, F. and {Pakmor}, R. and {Nelson}, D. and 
	{Genel}, S. and {Vogelsberger}, M. and {Naiman}, J. and {Torrey}, P.
	},
    title = "{Simulating galaxy formation with black hole driven thermal and kinetic feedback}",
  journal = {\mnras},
archivePrefix = "arXiv",
   eprint = {1607.03486},
     year = 2017,
    month = mar,
   volume = 465,
    pages = {3291-3308},
      doi = {10.1093/mnras/stw2944},
   adsurl = {http://adsabs.harvard.edu/abs/2017MNRAS.465.3291W}
}

@ARTICLE{pillepich18,
   author = {{Pillepich}, A. and {Springel}, V. and {Nelson}, D. and {Genel}, S. and 
	{Naiman}, J. and {Pakmor}, R. and {Hernquist}, L. and {Torrey}, P. and 
	{Vogelsberger}, M. and {Weinberger}, R. and {Marinacci}, F.},
    title = "{Simulating galaxy formation with the IllustrisTNG model}",
  journal = {\mnras},
archivePrefix = "arXiv",
   eprint = {1703.02970},
     year = 2018,
    month = jan,
   volume = 473,
    pages = {4077-4106},
      doi = {10.1093/mnras/stx2656},
   adsurl = {http://adsabs.harvard.edu/abs/2018MNRAS.473.4077P}
}

@ARTICLE{springel03,
       author = {{Springel}, Volker and {Hernquist}, Lars},
        title = "{Cosmological smoothed particle hydrodynamics simulations: a hybrid multiphase model for star formation}",
      journal = {\mnras},
         year = 2003,
        month = feb,
       volume = {339},
       number = {2},
        pages = {289-311},
          doi = {10.1046/j.1365-8711.2003.06206.x},
archivePrefix = {arXiv},
       eprint = {astro-ph/0206393},
 primaryClass = {astro-ph},
       adsurl = {https://ui.adsabs.harvard.edu/abs/2003MNRAS.339..289S}
}

\end{document}